\documentclass[bj, preprint]{imsart}

\usepackage[usenames,dvipsnames]{color}
\usepackage{amssymb,  amsfonts,color,latexsym}
\usepackage{amsmath}
\usepackage[numbers]{natbib}
\usepackage{bbm}

\newtheorem{thrm}{Theorem}
\newtheorem{prop}{Proposition}
 \newtheorem{lemma}{Lemma}
 \newtheorem{cor}{Corollary}
 
 \newtheorem{exa}{Example}

\usepackage{color, graphics}
\usepackage{bm}
\usepackage{graphicx}
\usepackage{subfigure}
\usepackage{pifont}
\usepackage{verbatim}
\usepackage{hyperref}

\newcommand{\Polya}{P\'{o}lya }
\newcommand{\bx}{\mathbf{x}}

\newcommand{\by}{\mathbf{y}}
\newcommand{\bz}{\mathbf{z}}
\newcommand{\bw}{\mathbf{w}}
\newcommand{\bu}{\mathbf{u}}
\newcommand{\bv}{\mathbf{v}}
\newcommand{\bV}{\mathbf{V}}
\newcommand{\bX}{\mathbf{X}}

\newcommand{\bmm}{\mathbf{m}}
\newcommand{\balpha}{{\mathbf{\alpha}}}
\newcommand{\bbeta}{{\mathbf{\beta}}}

\newcommand{\TT}{\mathbf{T}}

\newcommand{\tp}{\tilde{p}}
\newcommand{\tP}{\tilde{P}}

\begin{document}

\begin{frontmatter}

\title{Predictive Characterization of Mixtures of Markov Chains}
\runtitle{Predictive Characterization of Mixtures of Markov Chains}

\begin{aug}
  \author{\fnms{Sandra}  \snm{Fortini}\thanksref{a,e1}\ead[label=e1,mark]{sandra.fortini@unibocconi.it}},
  \author{\fnms{Sonia} \snm{Petrone}\thanksref{a,e2}\ead[label=e2,mark]{sonia.petrone@unibocconi.it}}
 
  \runauthor{S.Fortini, S. Petrone}

  \affiliation{Bocconi University, Milano}

  \address[a]{Department of Decision Sciences, Bocconi University, Via Roentgen 1, 20136 Milano, Italy. \printead{e1,e2}}

\end{aug}

\begin{abstract}
Predictive constructions are a powerful way of characterizing the probability laws of stochastic processes with
certain  forms of invariance, such as exchangeability or Markov exchangeability. When de Finetti-like
representation theorems are available, the predictive characterization implicitly defines the prior
distribution, starting from assumptions on the observables; moreover, it often helps in designing efficient
computational strategies. In this paper we give necessary and sufficient conditions on the sequence of
predictive distributions such that they characterize a Markov exchangeable probability law for a discrete valued
process $\bX$. Under recurrence, Markov exchangeable processes are mixtures of Markov chains. 
Our predictive conditions  are in some sense minimal sufficient conditions for Markov exchangeability; we also
provide predictive conditions for recurrence. We illustrate their application in relevant examples from the literature and in novel constructions. 
\end{abstract}


\end{frontmatter}

\section{Introduction}  \label{sec:intro}

Predictive characterization of the probability law of a stochastic process is a fundamental problem in probability and statistics.
Informally, this means characterizing the probability law $P$ of a process $(X_n, n \geq 1)$ through the sequence of predictive distributions $(P_n, n \geq 1)$, such that $X_1$ has distribution $P_1$ and $X_{n+1} \mid X_1, \ldots, X_n$ has distribution $P_n$ for $n \geq 1$.
The sequence $(P_n)$ characterizes a probability measure $P$ for the  stochastic process $(X_n)$ under general assumptions, by the Ionescu-Tulcea theorem. The problem of interest is to determine  the conditions under which it characterizes a  law $P$ with some given properties, and in particular, some specific invariance property. Necessary and sufficient conditions under which the sequence of predictive distributions
$(P_n)$ characterizes an exchangeable $P$ are given in \citep{fortini2000}. In the present paper, we give necessary and sufficient conditions for the sequence $(P_n)$ to characterize a $P$ which is partially exchangeable in the sense of Diaconis and Freedman \citep{diaconisFreedman1980} -- or, using the terminology of Zaman \citep{zaman1984} and Zabell \citep{zabell1995}, Markov exchangeable.

In Bayesian statistics, predictive characterizations have fundamental and practical relevance. Prediction is often the main goal of statistical analysis and, from a Bayesian perspective, the predictive approach seems  natural. Even in the context of independent replicates of an experiment,
probabilistic dependence is introduced through the assumption of exchangeability and prediction is naturally solved through the conditional distributions of future results given the observed facts.
Indeed, according to de Finetti \citep{definetti1937}, a statistical model is just a link of the probabilistic chain that leads from past to future events. Thus, at least in principle, models and priors on non-observable parameters can and should be induced by probability assertions on the observable $X_n$, such as exchangeability and predictive structures. The predictive characterization of prior distributions is a long studied problem in Bayesian statistics. Dirichlet conjugate priors for exchangeable categorical sequences have been characterized by Zabell \citep{zabell1982} based on Johnson's sufficiency postulate. Diaconis and Ylvisaker  \citep{diaconisYlvisaker1979} characterize  conjugate priors for the natural exponential family through predictive conditions. Powerful predictive constructions have also been given in Bayesian nonparametrics. The predictive characterization of the Dirichlet process in terms of \Polya sequences \citep{blackwellMcQueen1973} clarifies the relationship with random partitions in combinatorics and population biology \citep{ewens1972}. Walker and Muliere \citep{walkerMuliere1999} characterize neutral to the right processes through an extension of Johnson's sufficiency postulate. 
The general class of species sampling priors \citep{Pitman96}, which includes  the Dirichlet process and the two parameter Poisson--Dirichlet process \citep{permanPitmanYor1992}, is characterized in terms of the predictive distributions. Zabell \citep{zabell1995} extends the characterization of Dirichlet conjugate priors through Johnson's sufficiency postulate  to Markov exchangeable sequences. 
Reinforced processes (\cite{coppersmithDiaconis1987},  \cite{pemantle88, pemantle2007}) play an important role in  predictive constructions of exchangeable and Markov exchangeable sequences; references include 
\cite{walkerMuliere1997}, \cite{muliere2000}.  Diaconis and Rolles \citep{diaconisRolles2006} provide a predictive characterization of  conjugate priors for the transition matrix of a reversible Markov chain through edge reinforced random walks on a graph. Developments for variable order reversible Markov chains  are found in \cite{bacallado2011} and \cite{bacalladoFavaroTrippa2013}. Connections with the theory of vertex--reinforced jump processes are studied in  \cite{sabotTarres2014}.
Beyond the foundational issues, predictive characterizations are powerful tools in hierarchical modeling of symmetry structures and as generating algorithms which can be exploited for computational purposes, as recent developments at the interface between statistics and machine learning show. See for example the predictive construction of Markov exchangeable processes with countable unknown state space by Beal, Ghahramani and Rasmussen \citep{beal2002} (see also \cite{teh2006} and \cite{fortiniPetrone2012}), which has a wide application in hierarchical clustering and infinite hidden Markov models; or the  
predictive construction of the Indian buffet process  \citep{griffithGhahramani2005} for latent features allocation, whose de Finetti-like representation has been later provided in  \cite{thibauxJordan2007}. Refer to \cite{tehJordan2010} for an overview and further references.

\medskip

We recall the main concepts and provide some preliminary results in Section \ref{sec:preliminary}. We first review two characterizations of mixtures of Markov chains, in terms of Markov exchangeability of the process and of partial exchangeability of the matrix of successor states. 
The main point of this section is to revisit main basic results in a predictive approach and relate the prior to the predictive distributions. For exchangeable sequences, the prior measure is the limit law of the  sequence of predictive distributions; we formalize an analogous result for recurrent Markov exchangeable sequences in subsection \ref{sec:predProp}. This result further enhances the interest for predictive constructions and, therefore, for sufficient predictive conditions for Markov exchangeability.

Addressing the latter question is the main theoretical contribution of the paper, presented in Section \ref{sec:pred}.   We give necessary and sufficient conditions on the sequence of predictive rules $(P_n)$ under which they characterize a Markov exchangeable $P$ for a discrete valued process $(X_n, n \geq 1)$. Furthermore, we give some predictive  conditions for recurrence in Section \ref{sec:recurrence}. Under recurrence, a Markov exchangeable sequence is a
mixture of recurrent Markov chains and the predictive structure characterizes the mixing distribution; that is, in a subjective, Bayesian approach, the prior distribution on the unknown transition matrix. Thus, our results are useful for verifying if a predictive scheme characterizes a prior for Bayesian inference on Markov chains.

In Section \ref{sec:examples}, we illustrate the results through a novel predictive construction, defined as an edge reinforced random walk on a colored graph. To some extent, the proposed scheme is a generalization of the edge reinforced random walks  of \cite{diaconisRolles2006}, 
 where  colors allow the reinforcement of edges even when they are not crossed. 
The motivating idea is to introduce forms of probabilistic dependence or constraints on the random transition matrix through the reinforcement of both edges and colors.  The predictive conditions given in the previous sections are used to establish when the  proposed predictive scheme characterizes a Markov exchangeable process. Several proposals in the literature can be recovered as special cases of colored edge reinforced random walks, which, in this sense,  offer a unifying framework. 
 An advantage of the predictive approach is that it encourages prior distributions that are closed under sampling, as we illustrate in  subsection \ref{sec:partitionColors}. We provide examples of colored edge reinforced random walks for which the prior distribution can incorporate information and constraints on the transition matrix, and can be easily updated given the data. Finally, we discuss extensions through the introduction of latent variables which may complicate the analytic computations, but can be easily simulated. 
 
Some complements to these results and detailed proofs are provided in the Appendix.

\section{Overview and preliminary results} \label{sec:preliminary}

\subsection{Markov exchangeability} \label{sec:ME}
Let $S$ be a finite or countable set, containing at least two elements, and $\bX=(X_n,n\geq 0)$ be a discrete-time stochastic process taking values in $S$ and starting at a specific state $x_0\in S$. The process $\bX$ is a mixture of Markov chains if there exists a probability law $\mu$ on the set ${\cal P}$ of stochastic matrices on $S \times S$ endowed with the topology of element-wise convergence and the corresponding Borel sigma-algebra, such that,  for every $\bx=(x_1,\dots,x_n)$,
\begin{equation} \label{eq:mixtureMC}
p(\bx)=\int_{{\cal P}}\prod_{k=1}^n\tp_{x_{k-1},x_k} \, \mu(d\tp) 
\end{equation}
where $p(\bx)=P(\bX_{1:n}=\bx)$ with $\bX_{1:n}=(X_1,\dots,X_n)$. Equivalently, $\bX$ is a mixture of Markov chains if there exists a ${\cal P}$-valued random element $\tP$ such that, conditionally on $\tP$, $\bX$ is a Markov chain with transition matrix $\tP$. In Bayesian inference for Markov chains, the probability law $\mu(\cdot)$ plays the role of the prior distribution on the unknown transition matrix. 

In a predictive approach, interest is in understanding what probabilistic assumptions on the observable process $\bX$ imply (\ref{eq:mixtureMC}), that is, the existence of a prior distribution on the non-observable matrix $\tP$.
A main result by Diaconis and Freedman \citep{diaconisFreedman1980} shows that, for recurrent processes, such assumption is  Markov exchangeability.  
Recurrence here means that $P(X_n=x_0\; \mbox{infinitely often})=1$.  Markov exchangeability is defined as invariance under a certain kind of symmetry. 
Two finite strings $\bz$ and $\bz'$ are equivalent, written $\bz\sim \bz'$, if $\bz$ and $\bz'$ have the same first element and exhibit the
same number of transitions from $i$ to $j$, for every pair of states $i$ and $j$. The process $\bX$ is Markov exchangeable if $(x_0, \bx) \sim (x_0, \bx')$  implies $p(\bx)=p(\bx')$. 
It can be proved (\cite{diaconisFreedman1980}, Lemma 5) that, if $\bz'\sim\bz$, then $\bz$ and $\bz'$ have the same length, end at the same state and visit each state the same number of times. Hence, Markov exchangeability can been seen as invariance with respect to a class of permutations, namely those permutations that do not alter the number of transitions between any two states. This motivates the original term of {\em partial exchangeability} used by Diaconis and Freedman.   

 Markov exchangeability is a necessary condition for the process to be  representable as a mixture of Markov chains; under recurrence, it is also sufficient. 
The proof of this result is based on the exchangeability of the sequence of $x_0$-blocks. A $x_0$-block is defined as a finite sequence of states that begins at $x_0$ and contains no further  instances of $x_0$. Recurrence of the process $\bX$ ensures that the sequence of $x_0$-blocks is infinite. Under recurrence, Markov exchangeability implies exchangeability of
the sequence of $x_0$-blocks. Because a Markov exchangeable sequence with independent and identically distributed $x_0$-blocks is a Markov chain, it follows that $\bX$ is a mixture of Markov chains (\cite{diaconisFreedman1980}, Proposition 15 and Theorem 7). Thus, a recurrent Markov exchangeable process is a mixture of recurrent Markov chains.

An important point is that, if $\bX$ is a mixture of recurrent Markov chains for some specified $\mu$, one can determine $\mu$ from $\bX$. Let
$\TT_{i,j}(x_0,\bx)$ denote the number of transitions from $i$ to $j$ in the finite sequence $(x_0,\bx)$, $\TT(x_0,\bx)=[\TT_{i,j}(x_0,\bx)]_{i,j\in S}$ be the matrix of transition counts, 
$\TT_{i}(x_0,\bx)$ be the $i$th row of $\TT(x_0,\bx)$ and 
 $\TT_{i,\cdot}(x_0,\bx)= \sum_j \TT_{i,j}(x_0,\bx)$. 
Define $\hat{\TT}_{i,j}(x_0,\bx) = \TT_{i,j}(x_0,\bx)/ \TT_{i, \cdot}(x_0,\bx)$ if $\TT_{i,\cdot}(x_0,\bx) >0$ and
$\hat{\TT}_{i,j}(x_0,\bx)=\TT_{i,j}(x_0,\bx)=0$ if $\TT_{i, \cdot}(x_0,\bx)=0$. 
Then $\hat{\TT}_{i,j}(x_0,\bX_{1:n})$ converges to $\tP_{i,j}$ almost surely (a.s.)  as $n \rightarrow \infty$
(\cite{diaconisFreedman1980}, Section 4).  The limit matrix $\tP$ may not be a stochastic matrix on $S \times
S$, as the sum of the elements in a row  corresponding to a state that is not visited is zero. In fact, $\tP$ is
a stochastic matrix on the random set of the visited states, say $A_{\tP}$. In order for the limit matrix $\tP$ to
be a stochastic matrix (almost surely), we can use a conventional enlargement of the state space, as in
\cite{fortini2002}. Let us introduce an additional state $\partial$ and denote by $S^*$ the enlarged space $S
\cup \{\partial\}$. Then set $\hat{\TT}_{i,\partial}=1- \sum_{j \in S} \hat{\TT}_{i,j}$ and
$\hat{\TT}_{\partial, \partial}=1$. The enlarged matrix $[\hat{\TT}_{i,j}]_{i,j \in S^*}$ converges pointwise
almost surely to a stochastic matrix $\tP$ on $S^* \times S^*$. The mixing measure $\mu$ in (\ref{eq:mixtureMC}) is
uniquely determined as the probability law of $\tP$. We have understood that the probability measure $P$ on
$S^\infty$ has been extended to $(S^*)^\infty$, and $\cal P$ represents  the class of stochastic matrices on
$S^*$. For the sake of simplicity, we keep the same notation, and keep denoting by $\bX$ the 
coordinate process on $(S^*)^\infty$; the distinction  is clear from the context.

\subsection{Partial exchangeability of successor states}
A different characterization of mixtures of recurrent Markov chains, hinted in \cite{deFinetti1959} and \cite{zabell1995}, is developed by Fortini {\em et al.} \citep{fortini2002}, in terms of partial exchangeability of the matrix of successor states.
 The $n$th successor of a state $i \in S$ is  the state that follows the $n$th visit to $i$. More formally, for every $i\in S^*$, let $\tau_n(i)$ be the time of the $n$th visit to $i$, with the proviso that
$\tau_n(i)=\infty$ if state $i$ is not visited $n$ times. The $n$th successor state of $i$ is defined as 
$V_{i,n}=X_{\tau_n(i)+1}$ if $\tau_n(i)<\infty$ and $V_{i,n}=\partial$ otherwise. The successor matrix associated to $\bX$ is then defined as the array $\bV=[V_{i,n},i\in S^*,n\geq 1]$. If the process $\bX$ is recurrent and Markov exchangeable, then it is also strongly recurrent
\citep{fortini2002}; that is, if a state $i$ is visited, it is visited infinitely often, almost surely. Thus,
the rows of the matrix $\bV$ are infinite sequences, the $i$th row being a sequence in $S^\infty$ if $i$ is
visited, or a sequence equal to $(\partial, \partial, \ldots)$ if $i$ is not visited, or if $i=\partial$. Fortini {\em et al.}   \citep{fortini2002} show that the process $\bX$ is recurrent and Markov exchangeable if and only if the successor matrix $\bV$ is partially exchangeable by rows, in the sense of de Finetti, that is, there exists a stochastic matrix $\tP$ on $S^*$ such that, conditionally on $\tP$, the random variables
$(V_{i,n},i\in S^*,n\geq 1)$ are independent  with $P(V_{i,n}=j|\tP)=\tP_{i,j}$. By the properties of partially
exchangeable sequences, the random matrix $\tP$ is determined as
\begin{equation} \label{eq:empV}
\tP_{i,j}=\lim_{n\rightarrow\infty}\frac{1}{n}\sum_{k=1}^n\delta_{V_{i,k}}(\{j\}) \quad a.s.
\end{equation}
where  $\delta$ is the degenerate measure defined by $\delta_a(\{a\})=1$.
This gives an interpretation of the prior $\mu$ in (\ref{eq:mixtureMC}) as the limit law of the empirical distributions of the successor states.  
Another characterization can be given in terms of predictive distributions, as we discuss in the next subsection.

\subsection{Predictive properties} \label{sec:predProp}

The directing measure of an infinite exchangeable sequence can be characterized as the limit of the sequence of the 
predictive distributions $P(X_{n+1} \in \cdot \mid X_1, \ldots, X_n)$, which converges weakly almost surely to a random distribution $F$ such that the $X_i \mid F$ are a random sample from $F$; see \cite{aldous1985}, page 60.  Similar properties hold for recurrent Markov exchangeable processes. In this case, from the characterization  in terms of partial exchangeability of the successor matrix $\bV$,  discussed in the previous section, it follows that, almost surely
\begin{eqnarray*}
&& \lim_{n \rightarrow \infty} P(V_{i,n}=j \mid V_{i,1}, \ldots, V_{i,n-1}) \\
&=& \lim_{n \rightarrow \infty} P(V_{i,n}=j \mid V_{i,1}, \ldots, V_{i,n-1}, V_{l,k}, k=1, 2, \ldots, l \in S, l \neq i) = \tP_{i,j} , \quad j \in S.
\end{eqnarray*}
The predictive rules above refer to the successors states. In terms of the sequence $\bX$, the following result based on stopping times holds. Recall that $\tau_n(i)$ is the time of the $n$th visit to state $i$, and let ${\cal F}_{\tau_n(i)}$ be the sigma-algebra of
the events until the $n$th visit of $\bX$ to state $i$. Let $X_{\tau_n(i)+1}=X_{k+1}$ if $\tau_n(i)=k$, and $X_{\tau_n(i)+1}=\partial$ if $\tau_n(i)=\infty$.

\begin{thrm} \label{conv}
Let $\bX$ be a mixture of recurrent Markov chains with random transition matrix $\tP$. Then, for every $i\in
S^*$,
$$
\lim_{n\rightarrow\infty}P(X_{\tau_n(i)+1}=j|{\cal F}_{\tau_n(i)})
=\tP_{i,j} \quad \; \mbox{a.s.}
$$
\end{thrm}

\noindent {\sc Proof.} 
The result is immediate for $i=\partial$. Consider $i \in S$.
Denote by ${\cal F}_{\tP}$ the sigma-algebra generated by $\tP$, and by ${\cal F}_{\tau_n(i)}\vee {\cal F}_{\tP}$
the sigma algebra generated by ${\cal F}_{\tau_n(i)}\cup {\cal F}_{\tP}$. Then 
$$
P(X_{\tau_n(i)+1}=j|{\cal F}_{\tau_n(i)})=P(V_{i,n}=j|{\cal F}_{\tau_n(i)})= 
E(P(V_{i,n}=j|{\cal F}_{\tau_n(i)}\vee {\cal F}_{\tP} )|{\cal F}_{\tau_n(i)}).
$$
Partial exchangeability of the successors matrix $\bV$ implies that $V_{i,n}$ is conditionally independent of the other
successor states $(V_{i,k},k<n;V_{j,l},j\in S^*,j\neq i,l\geq 1)$, given $\tP$. As formally proved in Lemma \ref{check} in the Appendix, 
${\cal F}_{\tau_n(i)}$ is included in the  sigma algebra  generated by 
$(V_{i,k},k<n;V_{j,l},j\in S^*,j\neq
i,l\geq 1)$;  therefore, 
$V_{i,n}$ is also conditionally independent of ${\cal F}_{\tau_{n}(i)}$, given $\tP$. Hence 
$$
P(X_{\tau_n(i)+1}=j|{\cal F}_{\tau_n(i)}) = E( \tP_{i,j} \mid {\cal F}_{\tau_n(i)}) , 
$$
which converges to $E(\tP_{i,j}|{\cal F}_{\tau_\infty(i)})$, where ${\cal F}_{\tau_\infty(i)}=\vee_n {\cal
F}_{\tau_n(i)}$ is the sigma-algebra generated by $\cup_{n=1}^\infty {\cal F}_{\tau_n(i)}$. Since $V_{i,k}$ is measurable with respect to ${\cal F}_{\tau_\infty(i)}$ for all $k \geq 1$, then
$\tP_{i,j}=\lim_{n\rightarrow \infty}\frac{1}{n}\sum_{k=1}^n\delta_{V_{i,k}}
(\{j\})$ is ${\cal
F}_{\tau_\infty(i)}$-measurable, too. Thus, $ E(\tP_{i,j}|{\cal F}_{\tau_\infty(i)})=\tP_{i,j}$ almost surely and the proof is
complete.
\hfill $\diamond$

\medskip
A mixture of recurrent Markov chains has a random transition matrix with independent rows if and only if the probability of observing a transition from $i$ to $j$ depends only on  the past transition from state $i$. 

\begin{cor} \label{cor_indep} 
Let $\bX$ be a mixture of recurrent Markov chains with random transition matrix $\tP$. The rows of $\tP$  are stochastically independent if and only if, for all $n \geq 1$, $X_{n+1}$ is conditionally independent of 
$\bX_{1:n}$ given $(X_n, \TT_{X_n}(x_0,\bX_{1:n}))$.
\end{cor}
Although this fact appears to be known in the literature (hints are in \cite{zabell1995}; see also, for example, \cite{muliere2000}, \cite{beal2002}), for  completeness we provide a proof in Appendix (Corollary \ref{cor_indep2}),  based on Theorem \ref{conv}.
  Informally, the result follows from partial exchangeability of the matrix $\bV$ of successors states. For any $i$ and $j$, the probability of a transition to $j$, given past observations ending at $i$, 
is equal to the probability that $V_{i, n_{i}+1}=j$ conditionally on the sequence of successors, where  $n_{i}$ is the number of past successors of state $i$. If  such probability
only depends on $i$ and on the transitions from $i$, the probability distribution of  $V_{i,n_{i}+1}$, given the other successor states,  only depends on the successors of $i$; thus, the rows of $\bV$
are independent. By (\ref{eq:empV}), it follows that $\tP$ has independent rows. Conversely, if $\tP$ has
independent rows, then partial exchangeability by rows of $\bV$ reduces to independence and internal exchangeability of
the rows, therefore the probability of a transition from $i$ to $j$ only depends on transitions from $i$.

\section{Predictive characterization of Markov exchangeability} \label{sec:pred}

The basic question studied in this work is: when does a sequence of predictive rules characterize a Markov exchangeable process? 
Before stating the results, let us introduce some simplifying notation. We continue to denote finite sequences of elements in $S$ by bold letters, (e.g. $\bx,\by,\dots$), while non-bold letters (e.g. $x,y,\dots$) denote single elements of $S$. Unless otherwise specified, a string can coincide with the empty string, denoted by $\varnothing$. 
The predictive probabilities are denoted by $p(\by|x_0,\bx)$. Hence
$p(\by|x_0,\bx)=p(\bx,\by)/p(\bx)$ if $p(\bx)\neq 0$ and $p(\by|x_0,\bx)$ is defined arbitrarily if $p(\bx)=0$.
A string $(x_0,\bx,i)$ should be interpreted as a string of any length starting at $x_0$ and ending at $i$; this includes the string of length one: $(x_0)$, when $i=x_0$.

Markov exchangeability clearly implies the following predictive  properties: 

\begin{description}
\item
$A)$ $p(\by|x_0,\bx,i)=p(\by|x_0,\bx',i)$ for every $\by$ and every $i,\bx,\bx'$ such that
$(x_0,\bx,i)\sim (x_0,\bx',i)$;
\item
$B)$ $p(\by|x_0,\bx,i)=p(\by'|x_0,\bx,i)$ for every $i,\bx$ and every $\by,\by'$ such that $(i,\by)\sim (i,\by')$.
\end{description}

Condition $B)$ is sufficient for Markov exchangeability; in fact, condition $B)$ for $(x_0, \bx, i)=x_0$ is the definition of Markov exchangeability. Thus, verifying $B)$ is not easier than verifying Markov exchangeability directly from the joint distribution. Our aim is to show that a predictive condition weaker than $B)$ and a simpler form of condition $A)$ are jointly sufficient for Markov exchangeability. 

Condition $A)$ is the joint predictive sufficiency of the last state and the transition counts: for every $n\geq 1$, $(X_{n+1},X_{n+2},\dots )$ are conditionally 
independent of $\bX_{1:n}$, given $(X_n,\TT(x_0,\bX_{1:n}))$. This is because $\TT(x_0,\bx_{1:n})=\TT(x_0,\bx'_{1:n})$ and $x_n=x_n'$ imply $(x_0,x_1,\dots, x_n,x_{n+1},\dots,x_{n+k})\sim$ $(x_0,x_1',\dots, x_n',x_{n+1},\dots,x_{n+k})$. This property is the  analogous, for Markov exchangeable processes, of predictive sufficiency of the empirical distribution for exchangeable sequences. Clearly, a predictive sufficient statistic is also one-step-ahead predictive sufficient, that is, for any $n$,
$X_{n+1}$ is conditionally independent of $\bX_{1:n}$, given the statistic. Predictive sufficiency and one-step-ahead predictive sufficiency are different conditions, in general. But in the specific case where the statistic has the form $(X_n, \TT(x_0, \bX_{1:n}))$, they turn out to be equivalent.

\begin{prop} \label{one-step}
Let $\bX$ be an $S$-valued process starting at $x_0$. Then, for any $n$, $(X_n,\TT(x_0,\bX_{1:n}))$
is predictive sufficient if and only if it is one-step-ahead predictive sufficient.
\end{prop}

\noindent {\sc Proof.} In order to show the non-obvious implication, it is sufficient to write the conditional
distribution of $X_{n+1},\dots,X_{n+k}$, given $X_1,\dots,X_n$, as a product of conditional distributions and
notice that $(x_0,\bx_{1:n})\sim(x_0,\bx'_{1:n})$ implies $x_n=x'_n$ and 
$(x_0,x_1,\dots,x_n,x_{n+1},\dots,x_{n+j})\sim(x_0,x_1',\dots,x_n',x_{n+1},\dots,x_{n+j})$ for every
$j=1,\dots,k$.
\hfill $\diamond$
\medskip

Thus, we can replace condition $A)$ with one-step-ahead predictive sufficiency of 
$(X_n,\TT(x_0,\bX_{1:n}))$. However, such condition alone does not imply Markov exchangeability. As a counterexample, consider $S=\{0,1,2,\ldots\}$, $x_0=0$ and for $n \geq 1$, $p(j|x_0, x_1,\dots,x_{n-1})=1/n$ for $j=1,\dots, n$
and zero otherwise. Since $n=1+\sum_{i,j}\TT_{i,j}(x_0,x_1,\dots,x_{n-1})$, the vector $(X_n,\TT(x_0,\bX_{1:n}))$
is one-step-ahead predictive sufficient. However, $\bX$ is not Markov exchangeable. For example, $p(1,3,1,2,3)=0$ while $p(1,2,3,1,3)\neq 0$.
 
On the other hand, one-step-ahead predictive sufficiency of the transition counts, together with a weaker form  of condition $B)$, implies Markov exchangeability. This is shown in the following theorem. 
We denote by $\{\bx\}$ the set of distinct elements in $\bx$.

\begin{thrm} \label{pred}
Let $\bX=(X_n,n\geq 0)$  be an $S$-valued process such that $X_0=x_0$. Then  $\bX$ is Markov exchangeable if and only if both the following conditions hold: 
\begin{description}
\item[$a)$] $p(y|x_0,\bx,i)=p(y|x_0,\bx',i)$ for every $y\in {S}$ and every $i,\bx,\bx'$ such that $(x_0,\bx,i)\sim (x_0,\bx',i)$;

\item[$b)$] $p(\by|x_0,\bx,i)=p(\by'|x_0,\bx,i)$ for every $i$, $\bx$, $\by$ and $\by'$ such that $\by=(\bu,\bw,i,\bv,\bw,i)$
and $\by'=(\bv,\bw,i,\bu,\bw,i)$ with $\{i\},\{\bu\},\{\bv\},\{\bw\}$ disjoint.
\end{description}
\end{thrm}

The proof makes use of some lemmas, given below. Before proceeding, some remarks are in order. 

\medskip

\noindent {\bf Remark 1}. Under condition $a)$, condition $b)$ is equivalent to simultaneously having:
\begin{description}
\item[$ bi$)] $p(\bu,i,i|x_0,\bx,i)=p(i,\bu,i|x_0,\bx,i)$ for every $\bu,i$ with $\{i\},\{\bu\}$ disjoint;

\item[$bii$)] $p(\bu,i,\bv,i|x_0,\bx,i)=p(\bv,i,\bu,i|x_0,\bx,i)$ for every $\bu,\bv,i$ with $\{i\},\{\bu\},\{\bv\}$ disjoint.

\item[$biii$)] $p(\bu,j,\bw,i,\bv,j|x_0,\bx,i)=p(\bv,j,\bw,i,\bu,j|x_0,\bx,i)$ for every  $\bu,\bv,\bw,i,j$
with $\{i\}$,$\{\bu\}$, $\{\bv\}$,$\{j,\bw\}$ disjoint.
\end{description}
Conditions $ bi$) and $bii$) are obtained from b) by setting $\bw=\bv=\varnothing$ and $\bw=\varnothing$, respectively; $biii$) is recovered by  substituting $\bw$ with $(j,\bw)$ and canceling
$p(\bw|x_0,\bx,i,\bu,j,\bw,i,\bv,j)$ from the left-hand side and $p(\bw|x_0,\bx,i,\bv,j,\bw,i,\bu,j)$ from the
right-hand side. Similarly for showing that $bi)$--$biii)$ imply $b)$.

\medskip

\noindent{\bf Remark 2}. For a recurrent process $\bX$, when  conditions $a)$ and $b)$ of Theorem \ref{pred} hold for any $\bu,\bv, \bw$ not including $i$, they imply  exchangeability of the $i$-blocks, by Theorem 3.1 in \cite{fortini2000}. For a general process, if conditions  $bi)-biii)$ hold for any $\bu$, $\bv$ and $\bw$ not including $i$, they are equivalent to invariance under block-switch transformations, which in turn is equivalent to Markov exchangeability (\cite{diaconisFreedman1980}, Section 4). Theorem \ref{pred} says that, under condition $a)$, it 
is enough to check block-switch invariance on a subset of disjoint strings.

\medskip
Before proving Theorem \ref{pred}, we introduce some convenient notation. The length of a string $\bz$ is denoted by
$|\bz|$. Given two finite strings $\bz$ and $\bz'$, we say that $\bz$ is shorter than $\bz'$ and write $\bz
\preceq \bz'$, if there exists $\bx$ such that $(\bx,\bz)\sim \bz'$; if $\bx\neq \varnothing$, we say that $\bz$
is strictly shorter than $\bz'$ and write $\bz \prec \bz'$. Given a class $\cal C$  of non empty $S$-valued
finite strings, an element $\bz^*$ is called minimal in $\cal C$ if $\bz^*\in\cal C$ and there is no $\bz \in
\cal C$ that is strictly shorter than $\bz^*$.

\begin{lemma}
\label{ii} Let $\bx_0=(\bu_0,j,j,\bv_0,k,j)$ with $k\neq j$ and let ${\cal C}=\{\bx\preceq\bx_0:
\bx=(\bu,j,j,\bv,k,j)\}$. Then there exists a minimal element $\bx^*$ in
 ${\cal C}$, given by $\bx^*=(j,j,\bv^*,k,j)$ with $j\not\in \{\bv^*\}$.
\end{lemma}

\noindent {\sc Proof.} The lemma is proved by contradiction. If there was no minimal element, we could find an infinite sequence
$\bx_0,\bx_{1},\bx_{2},\dots$ of strings in $\cal C$ such that $\bx_0\succ \bx_{1}\succ\bx_{2}\succ\dots$. In
that case, $|\bx_0|> |\bx_{1}|>|\bx_{2}|>\dots$, which is impossible, since $|\bx_0|$ is finite. Thus, a minimal
element exists. Let such minimal element be $\bx^*=(\bu^*,j,j,\bv^*,k,j)$. Then, $\bu^*$ has to be empty, as
otherwise a shorter element of $\cal C$ could be obtained by deleting $\bu^*$. Furthermore, should $\bv^*$
contain $j$, we would have $\bx^*=(j,j,\bv^*_1,j,\bv^*_2,k,j)\sim (j,\bv^*_1,j,j,\bv^*_2,k,j)$ and a shorter
string could be obtained by deleting $(j,\bv^*_1)$. \hfill$\diamond$

\begin{lemma}
\label{lem2} Let $j,k,k'$ be distinct elements of $S$ and let $\bx=(\bu,k,j,\bv,k',j)$ and
$\bx'=(\bu',k',j,\bv',k,j)$ be such that $\bx\sim\bx'$. Then, either $\{\bu,k\}\cap\{j,\bv,k'\}\neq\emptyset$ or
$\{\bu',k'\}\cap\{j,\bv',k\}\neq\emptyset$ or both.
\end{lemma}

\noindent {\sc Proof.} By contradiction: suppose that $j\not\in \{\bu\}$, $j\not\in \{\bu'\}$,
$\{\bu,k\}\cap\{\bv,k'\}=\emptyset$ and $\{\bu',k'\}\cap\{\bv',k\}=\emptyset$. Then $k\not\in\{\bv\}$ and
$k\not\in\{\bu'\}$. Let $\bu=(u_1,\dots,u_m)$. We show by backward induction on $s$ that $u_s\in \{\bv',k\}$ for
every $s=1,\dots,m$. $u_m$ is a predecessor of $k$ in $\bx$ and therefore in $\bx'$. Since $k\not\in\{\bu'\}$,
then $u_m\in \{\bv'\}$. Now, suppose that $u_s\in \{\bv'\}$. Since $u_{s-1}$ is a predecessor of $u_s$ in $\bx$,
it is also a predecessor of $u_s$ in $\bx'$. Since $u_s\not\in \{\bu'\}$, then $u_{s-1}\in\{\bv'\}$. It follows
by induction that $u_s\in \{\bv'\}$ for every $s$. Since $u_1\in \{\bv'\}$ and $\{\bu'\}\cap\{\bv'\}=\emptyset$,
then $u_1\not\in \{\bu'\}$. Hence, the first element of $\bx'$ is not $u_1$, which contradicts $\bx\sim\bx'$.
\hfill $\diamond$

\begin{lemma}
\label{lem3} Let $j,k,k'$ be distinct elements of $S$ and let $\bx_0=(\bu_0,k,j,\bv_0,k',j)$ with
$\{\bu_0,k\}\cap \{j,\bv_0,k'\}\neq \emptyset$. Let ${\cal C}=\{\bx \preceq\bx_0:
\bx=(\bu,k,j,\bv,k',j),\{\bu,k\}\cap \{j,\bv,k'\}\neq \emptyset\}.$ Then,  there exists a minimal element
$\bx^*=(\bu^*,k,j,\bv^*,k',j)$ in
 ${\cal C}$ which is either of the form $\bx^*=(j,\bu,k,j,\bv,k',j)$, with $\{j\},\{k\},\{k'\},\{\bu\},\{\bv\}$
 disjoint, or $\bx^*=(i,\bu,k,j,\bw,i,\bv,k',j)$,
 with $\{i\},\{k\},\{k'\},\{\bu\},\{\bv\},\{j,\bw\}$ disjoint.

 \end{lemma}

The existence of a minimal element can be proved as for Lemma \ref{ii}. The rest of the proof is by
contradiction: if $\bx^*$ does not have the above structure, one can make a switch transformation by moving a
piece of $\bx^*$ in front, and, by deleting such a piece, obtain a string in ${\cal C}$ which is shorter than $\bx^*$. A detailed proof is provided in the Appendix.

\bigskip
\noindent {\sc Proof of Theorem \ref{pred}.}
Markov exchangeability implies conditions $a)$ and $b)$.
In order to prove that, under $a)$ and $b)$, $(x_0,\bz)\sim(x_0,\bz')$ implies $p(\bz)=p(\bz')$, we proceed by induction
on $n=|\bz|$. It is convenient to use the conditions 
$bi)$--$biii)$. 
The thesis is true for $n=1$. Suppose it is true for all sequences of length $k\leq n$ and let 
$\bz=(z_1,\dots,z_{n+1})$, $\bz'=(z'_1,\dots,z'_{n+1})$ be two strings such that $(x_0,\bz)\sim(x_0,\bz')$. Since equivalent strings end at the same state, $z_{n+1}=z'_{n+1}$. 
We treat separately the cases $z_n=z_n'$ and  $z_n\neq z_n'$.

If $z_n=z_n'$, then $(x_0,z_1,\dots,z_n)\sim (x_0,z_1',\dots,z_n')$. By the induction hypothesis,
$p(z_1,\dots,z_n)= p(z_1',\dots,z_n')$. Hence,
$$p(\bz)=p(z_{n+1}|x_0,z_1,\dots,z_n) p(z_1,\dots,z_n) =p(z'_{n+1}|x_0,z_1',\dots,z_n') p(z_1',\dots,z_n')=p(\bz')$$
where we have used $z_{n+1}=z'_{n+1}$  and condition $a)$.

Suppose, now, that $z_n\neq z_n'$. Let $z_n=k$, $z_n'=k'$ and $z_{n+1}=z_{n+1}'=j$. It is enough to
distinguish two cases: (I): $j=k'$ and (II): $j\neq k,k'$. \\
(I) Since $(x_0,\bz')$ contains the transition $(j,j)$ 
and $(x_0,\bz) \sim (x_0,\bz')$, 
$(x_0,\bz)$ also contains $(j,j)$. Hence we can write $(x_0,\bz)=(x_0,\bu_0,j,j,\bv_0,k,j)$ with $k\neq j$.
The last expression should be intended as a string starting at $x_0$, containing the transition $(j,j)$ in some
position and the transition $(k,j)$ in the last position. Such strings  include $(x_0,x_0,\bv_0,k,x_0)$, for $j=x_0$. By
Lemma \ref{ii}, there exist $\bx$ and $\bv^*$ such that $(x_0,\bz)\sim (x_0,\bx,j,j,\bv^*,k,j)$ with
$j\not\in\{\bv^*\}$. Let $\bu=(\bv^*,k)$. Since $z_n=k$, again as in the first part of the proof, $p(\bz)=p(\bx, j,j, \bv^*, k,j)$. Furthermore, $p(\bz)=p(\bx,j)p(j,\bu,j|x_0,\bx,j)$ and, by condition $bi)$,
with $i=j$, $p(\bx,j)p(j,\bu,j|x_0,\bx,j)=p(\bx,j)p(\bu,j,j|x_0,\bx,j)$. Hence $p(\bz)=p(\bx,j,\bu,j,j)$.
Since $z'_n=j$, reasoning as in the first part of the proof, $p(\bx,j,\bu,j,j)=p(\bz')$. Thus $p(\bz)=p(\bz')$.\\
(II) We can write $(x_0,\bz)=(\bu_0,k,j,\bv_0,k',j)$ and  $(x_0,\bz')=(\bu'_0,k',j,\bv'_0,k,j)$. By Lemma
\ref{lem2}, without loss of generality, we can suppose that $\{\bu_0,k\}\cap\{j,\bv_0,k'\}\neq\emptyset$. Let us consider
the class ${\cal C}=\{(\bu,k,j,\bv,k',j)\preceq(x_0,\bz) : \{\bu,k\}\cap \{j,\bv,k'\}\neq \emptyset\}$. By Lemma
\ref{lem3}, a minimal element in ${\cal C}$ exists and can be written either as $\bx^*=(j,\bu,k,j,\bv,k',j)$,
with $\{j\},\{k\},\{k'\},\{\bu\},\{\bv\}$
 disjoint, or
as $\bx^*=(i,\bu,k,j,\bw,i,\bv,k',j)$, with $\{i\}$,$\{k\}$,$\{k'\}$, $\{\bu\}$,$\{\bv\}$,$\{j,\bw\}$ disjoint. In the
first case, there exists $\bx$ such that $(x_0,\bz)\sim (x_0,\bx,j,\bu,k,j,\bv,k',j)$. Since $z_n=k'$, again 
as in the first part of the proof, $p(\bz)=p(\bx,j,\bu,k,j,\bv,k',j)$. Furthermore
$p(\bx,j,\bu,k,j,\bv,k',j)=p(\bx,j)p(\bu,k,j,\bv,k',j|x_0,\bx,j)$, which, by $bii)$ with $i=j$, is equal to 
$$p(\bx,j)p(\bv,k',j,\bu,k,j|x_0,\bx,j)=p(\bx,j,\bv,k',j,\bu,k,j). $$ Since $z'_n=k$, then 
$p(\bx,j,\bv,k',j,\bu,k,j)=p(\bz')$. If the minimal element is $\bx^*=(i,\bu,k,j,\bw,i,\bv,k',j)$ the proof can
be obtained along the same steps by using property $biii)$. \hfill $\diamond$.

\begin{exa} {\bf Edge reinforced random walks}. 
{\em Edge reinforced random walks (ERRW) are predictive schemes that characterize conjugate priors for reversible Markov chains
(\cite{coppersmithDiaconis1987}, \cite{rolles2003}, \cite{diaconisRolles2006}). Consider a finite undirected graph $G$ with vertex
set $V$ and edge set $E$ (possibly including loops). All edges in $E$ are given a strictly positive weight; at time zero, edge $e$ has weight $\alpha_e > 0$.  An
edge-reinforced random walk on $G$ with starting point $x_0 \in V$ is defined as follows. The process starts at
$x_0$ at time $0$. At each step, the random walker traverses an edge with probability proportional to its
weight. Each time an edge in $E$, that is not a loop, is traversed, its weight is increased by 1. Each time a
loop in $E$ is traversed, its weight is increased by 2. Thus the predictive probability of traversing edge 
$e=(i,j)$ is
\begin{equation} \label{eq:ERRWpred} 
p(j|x_0,\bx,i)=\frac{\alpha_{(i,j)}+\TT_{i,j}(x_0,\bx,i)+\TT_{j,i}(x_0,\bx,i)}
{\alpha_{(i,\cdot)}+\TT_{i,\cdot}(x_0,\bx,i)+\TT_{\cdot,i}(x_0,\bx,i)}
\end{equation}
where $\alpha_{(i,\cdot)}=\sum_k\alpha_{(i,k)}$ is the sum of the weights of the edges incident to $i$, and $\TT_{i,\cdot}=\sum_{j'}\TT_{i,j'}$ and $\TT_{\cdot,i}=\sum_{j'}\TT_{j',i}$ are the transitions from and to state $i$, respectively. ERRWs are known to generate Markov exchangeable processes. We 
notice that Markov exchangeability can be easily verified through Theorem \ref{pred}. Condition $a)$ is immediate, as (\ref{eq:ERRWpred}) depends on $(x_0,\bx,i)$ only through the transitions and
the last state $i$. Condition $b)$ is satisfied if $p(\by \mid x_0,\bx,i)=p(\by' \mid x_0,\bx,i)$ for vectors
$\by , \by'$ of the form $\by=(\bu, \bw, i, \bv, \bw,i)$ and $\by'=(\bv, \bw,i, \bu, \bw,i)$ with
$\{i\},\{\bu\}, \{\bv\}, \{\bw\}$ disjoint. Direct computation of the above conditional probabilities involves
the product of terms of the form (\ref{eq:ERRWpred}) recursively updated. The transition counts satisfy
$$ \TT_{u_{l-1},u_{l}}(x_0,\bx, i, \bv, \bw, i, u_1, \ldots, u_{l-1})=
\TT_{u_{l-1},u_{l}}(x_0,\bx, i, u_1, \ldots, u_{l-1})
$$
and similarly for $ \TT_{u_{l},u_{l-1}}$ and for all the transition counts in the numerator of (\ref{eq:ERRWpred}). Analogous equations hold for the denominators, with the exception of the terms involving $i$ and $\bw$, which  satisfy
$$\TT_{i,\cdot}(x_0,\bx, i, \bu, \bw, i)=\TT_{i,\cdot}(x_0,\bx, i, \bv, \bw, i)= \TT_{i,\cdot}(x_0,\bx, i)+1 , $$
and similarly for $\TT_{\cdot,i}(x_0,\bx,i)$ and $\bw$. The above equations imply condition b).}
\end{exa}

\medskip

If  $(X_n,\TT_{X_n}(x_0, \bX_{1:n}))$ is one-step-ahead predictive sufficient, that is, $p(j \mid x_0, \bx, i)$ is a function $\pi(j \mid \TT_{i}, i)$ of the last element $i$ and of the transition counts $\TT_i=\TT_i(x_0, \bX_{1:n})$ from $i$, then 
 the conditions for Markov exchangeability simplify greatly.

\begin{cor} \label{tindip}
Let $\bX$ be an $S$-valued stochastic process starting at $x_0$ and such that  $(X_n,\TT_{X_n})$ 
is one-step-ahead predictive sufficient. Then the process $\bX$ is Markov exchangeable if and only if, for every $i,u,v$,
\begin{equation} \label{indip}
\pi(u \mid \TT_{i}, i) \, \pi(v \mid \TT_{i} +{\bf e}_u, i) = \pi(v \mid \TT_{i}, i) \, \pi(u
\mid \TT_{i} + {\bf e}_v, i)
\end{equation}
where $\TT_{i}+{\bf e}_k$ is the vector $\TT_{i}$ with the $k$th element incremented by one.
\end{cor}

\noindent {\sc Proof}. 
The proof is simple, applying Theorem \ref{pred}.
Condition $a)$ holds by assumption.
We need to show that,  under the assumptions of the Corollary,
(\ref{indip}) is equivalent to condition $b)$ of Theorem \ref{pred}, that is, for any $i$, $\bu=(u_1,\bu')$,
$\bv=(v_1, \bv')$, ${\bw}=(w_1,\bw')$ disjoint,
\begin{equation} \label{eq:predu-v}
p(u_1, \bu', w_1,\bw', i, v_1, \bv', w_1,\bw' \mid x_0,\bx, i)= p(v_1, \bv', w_1,\bw', i, u_1, \bu', w_1,\bw'
\mid x_0,\bx, i).
\end{equation}
Writing both sides as products of conditional probabilities, and noticing that
\begin{align*}
&p(\bu' \mid x_0,\bx,i, u_1)=p(\bu' \mid x_0,\bx,i, \bv,  \bw, i, u_1); \quad
p(w_1\mid x_0,\bx,i,\bu)=p(w_1\mid x_0,\bx,i,\bv,\bw,i,\bu); \\
&p(\bw' \mid x_0,\bx, i, \bu,w_1)=p(\bw' \mid x_0,\bx, i, \bv, w_1);
\quad p(i \mid x_0,\bx, i, \bu,\bw)=p(i \mid x_0,\bx, i, \bv,\bw); \\
& p(\bv' \mid x_0,\bx, i, \bu, \bw, i, v_1)=p(\bv'\mid x_0,\bx, i, v_1) ;
\quad p(w_1 \mid x_0,\bx, i, \bu, \bw, i, \bv)=p(w_1 \mid x_0,\bx, i, \bv); \\
& p(\bw' \mid x_0,\bx, i, \bu, \bw, i, \bv,w_1)=p(\bw' \mid x_0,\bx, i, \bv, \bw,i,\bu,w_1), &
\end{align*}
by the predictive sufficiency assumptions, one can easily verify that (\ref{eq:predu-v}) holds if and
only if
$
p(u_1 \mid x_0,\bx, i)p(v_1 \mid x_0,\bx, i, \bu, \bw, i)= p(v_1 \mid x_0,\bx, i)p(u_1 \mid x_0,\bx, i, \bv,
\bw, i),
$
which, again by the predictive sufficiency assumptions, corresponds to (\ref{indip}). 
\hfill $\diamond$

\medskip

An alternative proof, which however requires the additional assumption that $\bX$ is recurrent, can be given in terms of the successor states.  If $\bX$ is recurrent, the matrix $\bV$ has rows of infinite length in $S$ and, under the assumptions of Corollary \ref{tindip}, such rows are independent. Then, condition (\ref{indip}) is equivalent to exchangeability of the sequence of successor states of $i$, by Theorem 3.1 in \cite{fortini2000}, for any $i$; therefore,  it is equivalent to Markov exchangeability, as shown in Section 2.

\begin{exa} \label{hoppe} {\bf Reinforced urn schemes}.
{\em
Let $S$ be finite or countable. Consider the following predictive probabilities 
\begin{equation} \label{eq:Hoppepred}
p(j\mid x_0,\bx,i)=\pi(j\mid \TT_{i}, i)=\frac{\alpha_{i} q_{i}(j)+\TT_{i,j}}{\alpha_{i}+\TT_{i, \cdot}}, 
\end{equation}
where $\sum_{j \in S} q_{i}(j)=1$ and $\alpha_i >0$ for any $i$.
The predictive probability of observing state $j$ is
a weighted average of an initial weight $q_{i}(j)$ and the relative transition counts from the last element in the sample, $\TT_{i,j}/\TT_{i, \cdot}$. 
 This is a simple example of a predictive structure as considered in Corollary \ref{tindip}.  It is immediate to verify that
 (\ref{indip}) holds; thus, by Corollary \ref{tindip}, the sequence of predictive distributions (\ref{eq:Hoppepred})
 characterizes a Markov exchangeable probability law for the process $\bX$.}

{\em For a finite state space $S$, the predictive rule (\ref{eq:Hoppepred}) has been derived by Zabell \citep{zabell1995} from Johnson's sufficiency postulate: he shows that,  if $\bX$ is recurrent and Markov exchangeable, and the predictive probability $p(j \mid x_0,\bx, i)$ is a function of $i, j, \TT_{i,j}(x_0,\bx, i), \TT_{i, \cdot}(x_0,\bx, i)$ for each $j$, then such function has to be of the linear form (\ref{eq:Hoppepred}). Here, the linear structure (\ref{eq:Hoppepred}) of the predictive probabilities is an assumption, while Markov exchangeability is deduced by Corollary \ref{tindip}.  
Muliere, Secchi and Walker \citep{muliere2000} construct the predictive rule (\ref{eq:Hoppepred}) 
through a reinforced urn process. An extension of their construction, for a countable state space $S$, can be obtained through a reinforced Hoppe's  urn scheme. 
A Hoppe urn \citep{hoppe1984} is associated to each state, with urn $i$ having initial number of black balls $\alpha_i$ and color distribution $q_i(\cdot)$ on $S$. The process starts at  $x_0$. At step $n$, a ball is drawn from urn $x_{n-1}$ and, if colored, it is returned in the urn together with an additional ball of the same color; if black, a color is picked from the color distribution $q_{x_{n-1}}(\cdot)$ and a ball of the sampled color is added in the urn, together with the black ball. The process then moves to the urn associated with the color of the additional ball, and so on.
$X_n$ represents the color of the additional ball at the $n$th step. The predictive law of the process $\bX$ defined in such way is (\ref{eq:Hoppepred}). If the process $\bX$ is recurrent,  then it is a mixture of recurrent Markov chains, whose random transition matrix has independent rows. From the results discussed in Section 2, the $i$th row is the random limit of the  sequence of predictive distributions of the successors of state $i$; here, 
the draws from the urn associated to state $i$. These form an exchangeable sequence with predictive rule (\ref{eq:Hoppepred}) which characterizes a Dirichlet prior distribution with parameters 
$(\alpha_i q_i(j), j \in S)$ when $S$ is finite, or a Dirichlet process with parameter $\alpha q_i(\cdot)$, denoted 
$DP(\alpha \, q_i(\cdot))$, if $S$ is countable. Extensions of the predictive rule (\ref{eq:Hoppepred}), introducing dependence across the rows of the transition matrix, are given in \cite{teh2006}, \cite{beal2002}, \cite{fortiniPetrone2012}, using hierarchical Hoppe's urns.
 
A mixture of Markov chains with random transition matrix having independent Dirichlet rows has predictive rule (\ref{eq:Hoppepred}).  
Thus, the Markov  exchangeable process $\bX$ characterized by (\ref{eq:Hoppepred}) can be represented as a mixture of Markov chains; but without recurrence such representation may not be unique.
We provide predictive conditions for recurrence in the next section.
}
\end{exa}

\section{Predictive conditions for  recurrence} \label{sec:recurrence}

A Markov exchangeable process $\bX$ is a mixture of processes of the following kinds (\cite{diaconisFreedman1980}, page 124): 1) recurrent Markov chains, 2) processes starting with a string
of transient states and continuing as recurrent Markov chains, 3) totally transient processes. It is of
interest to have conditions under which $\bX$ is a mixture of Markov chains, thus excluding mixing of processes of kind 2 and of processes of kind 3 which are not Markov chains. Recurrence is a sufficient condition for restricting to mixtures of processes of  kind 1. A recurrent process $\bX$ is Markov exchangeable if and only if it is a mixture of recurrent Markov chains. 
In this sense, recurrence is a simplifying assumption. 

Proving recurrence of a general Markov exchangeable process is a difficult task, and results are mostly available for specific constructions. 
In a predictive approach, interest is in conditions on the predictive distributions that imply recurrence. We provide below a sufficient predictive condition for recurrence, for a general process. For a
Markov exchangeable process, this condition is also necessary.

\begin{thrm} \label{th:rec}
If, for any $(x_1,x_2,\dots)$ in a set of probability one,
\begin{equation}\label{recsum}
\sum_{n=0}^\infty p(x_0|x_0,x_1,\dots,x_n)=\infty,
\end{equation}
then $\bX$ is recurrent. Conversely, if $\bX$ is Markov exchangeable and recurrent, then (\ref{recsum}) holds, almost surely.
\end{thrm}

\noindent {\sc Proof.} We first prove that (\ref{recsum}) implies recurrence. The proof is based on L\'evy's extension of the Borel Cantelli lemma. Let $A_n=\{X_n=x_0\}$ and ${\cal F}_n$ the sigma-algebra generated by $\bX_{1:n}$. Then $A_n\in {\cal F}_n$ and (\ref{recsum}) can be rewritten as $\sum_{k=1}^\infty P(A_k\mid {\cal F}_{k-1})=\infty$, almost surely. This implies (see e.g. \cite{williams1991}, Section 12.15)  
$$
\frac{\sum_{k=1}^n \mathbbm{1}_{A_k}}{\sum_{k=1}^n P(A_k| {\cal F}_{k-1})}\rightarrow 1 \quad  a.s.
$$ as $n\rightarrow\infty$, where $\mathbbm{1}_A$ is the indicator of event $A$.   
In turn, this entails $P(A_n\; \mbox{infinitely often})=1$. 

Now we prove the second assertion. If $\bX$ is recurrent and Markov exchangeable, it is a mixture of recurrent
Markov chains. The process $\bX$ is also strongly recurrent, that is, $P(X_n=i \; \mbox{infinitely often} \mid \mbox{$i$ is visited})=1$,
for any state $i$. Furthermore, the set of elementary events $(x_1,x_2,\dots)$ such that $\tp_{i,x_0}=0$ for every visited
$i$,  has probability zero as, otherwise, $x_0$ would not be visited infinitely many times with probability one.
It follows that, for any $(x_1,x_2,\dots)$ in a set of probability one, there exists a state $i$, generally
depending on $(x_1,x_2,\dots)$, that is visited infinitely many times, has positive transition probability 
$\tp_{i,x_0}$ and satisfies  $p(x_0 \mid x_0, \dots,x_{\tau_n(i)}) \rightarrow \tp_{i,x_0}
>0$, where the last assertion follows from Theorem \ref{conv}. Therefore,
$$
\sum_{n=0}^\infty p(x_0|x_0,\dots x_{n}) \geq \sum_{n=1}^\infty p(x_0|x_0,\dots,x_{\tau_n(i)})=\infty. 
$$
\hfill $\diamond$

\medskip

Theorem \ref{th:rec} offers a strategy to prove recurrence; in some cases, one can find a lower bound for $p(x_0|x_0, \dots, x_{n})$ and easily prove that the series diverges.

\medskip
\noindent {\bf Example \ref{hoppe} -- Ctd}.
Consider again the reinforced Hoppe's urns  of Example 
\ref{hoppe} 
and suppose that $\alpha_i=\alpha$ and $q_i(\cdot)=q(\cdot)$ for every $i$ (all urns have the same
initial number of black balls and the same color distribution), with $q(x_0)>0$.  In this case, recurrence is an immediate
consequence of Theorem \ref{th:rec}, since $\sum_{n=0}^\infty p(x_0|x_0, x_1,\dots,x_n) \geq \sum_{n=0}^\infty
\alpha q(x_0) / (\alpha+n)=\infty$. For this particular case, recurrence was verified in \cite{fortiniPetrone2012} through a different approach. 
 The general case (different $\alpha_i$ and $q_i$)
is still immediate if $\inf_i \alpha_i q_i(x_0)>0$. A simple example is obtained when the state space is finite and $q_i(x_0)>0$ for any $i$. 

\medskip
In the example above, if $q_i(x_0)=0$ for some state $i$, recurrence is more difficult to prove. In fact, if there exists some $i\in S$ that is visited infinitely many times almost surely and such that $q_i(x_0)>0$, we could still apply Theorem
\ref{th:rec}. However the existence of such $i$ depends precisely on the recurrence property that we
are exploring. The following theorem can be useful in this situation. Recall that ${\cal F}_{\tau_n(i)}$ is the sigma-field generated by all  events until the $n$th visit of $\bX$ to state $i$.

\begin{thrm}
\label{rec2} Given $i, j \in S$, if
\begin{equation} \label{eqrec2}
\sum_{n=1}^\infty [P(X_{\tau_n(i)+1}=j|{\cal F}_{\tau_n(i)}) 
\mathbbm{1}_{\{\tau_n(i)< \infty\}} +
\mathbbm{1}_{\{\tau_n(i)=\infty\}}]=\infty
\end{equation}
almost surely, then $P( (X_n=i \; \mbox{for finitely many $n$}) \cup (X_n=j \; \mbox{infinitely often}))=1$.
\end{thrm}

\noindent {\sc Proof.} For every $n\geq 1$, let $A_{n+1}=(\tau_{n}(i)=+\infty)\cup (X_{\tau_{n}(i)+1}=j)$ and ${\cal G}_n={\cal F}_{\tau_n(i)}$. 
Then, $A_n \in {\cal G}_{n}$, and (\ref{eqrec2}) can be written as 
$
\sum_{n=1}^\infty P(A_{n+1}|{\cal G}_{n}) =\infty, 
$
almost surely. By L\'evy's extension of the Borel-Cantelli lemma,  
$$ 
\frac{\sum_{k=2}^n \mathbbm{1}_{A_k}}{\sum_{k=2}^n P(A_k\mid {\cal G}_{k-1})}\rightarrow 1
$$
almost surely, as $n\rightarrow\infty$. This implies that $\sum_{k=2}^\infty \mathbbm{1}_{A_k}=\infty$ a.s. The thesis follows.
\hfill $\diamond$

\medskip

Theorem \ref{rec2} implies that, if there exists a state $i$ that is visited infinitely often, it is sufficient to check that (\ref{eqrec2}) holds with $j=x_0$ to have recurrence. 
\medskip

\noindent{\bf Example \ref{hoppe} -- Ctd}. 
Consider the urn scheme of Example \ref{hoppe} with general weights $\alpha_i  q_i(\cdot)$. 
Let $Q$ denote the stochastic matrix with $(i,j)$th entry $Q_{i,j}=q_i(j)$. Consider two states $i$ and $j$ such that $q_i(j)>0$.
Then, almost surely,
\begin{eqnarray*}
\sum_{n=1}^\infty [P(X_{\tau_n(i)+1}=j \mid 
{\cal F}_{\tau_n(i)}) \mathbbm{1}_{\{\tau_n(i)< \infty\}} +
\mathbbm{1}_{\{\tau_n(i)=\infty\}}]
\!\geq \!\sum_{n=1}^\infty [\frac{\alpha_i q_i(j)}{\alpha_i+n} \mathbbm{1}_{(\tau_n(i)<\infty)} +
\mathbbm{1}_{(\tau_n(i)=\infty)} ]=\infty .
\end{eqnarray*}
Hence, by Theorem \ref{rec2}, if the chain visits $i$ infinitely many times, it also visits $j$ infinitely many
times. In the case when $q_i(j)=0$, if state $j$ can be reached from $i$ in a finite number of steps, that is, there exist $u_1, \ldots, u_k$
such that $Q_{i,u_1} \cdots Q_{u_k,j} >0$, we can use the same reasoning to show that condition
(\ref{eqrec2}) holds for each pair of those states. Thus, if $j$ is accessible from $i$ and $i$ is recurrent,
then $j$ is recurrent as well. 
A recurrent state $i$ certainly exists if the state space is finite. Thus, in  the finite case, we can give a simple sufficient condition for recurrence: if the matrix $Q$ is the transition matrix of an irreducible Markov chain, then $\bX$ is recurrent. 

\medskip

We should notice that these techniques are mostly useful for processes with a finite state space, where one can assume that at least one state is recurrent. Although they can also be helpful in fairly simple constructions with countable state space, such as in Example 2, proving recurrence in the infinite case is generally far more difficult than in the finite case and requires more sophisticated techniques.
	
Recurrence has been studied in depth for edge reinforced random walks. 
A phase transition in the recurrence and transience of ERRWs on an infinite binary tree has been shown by Pemantle \citep{pemantle1988b}. 
For ERRWs on {\em finite} graphs, Keane and Rolles (\cite{keaneRolles2000}, Proposition 1) proved recurrence using techniques similar to those used here for proving Theorems 1 and 2. Exploiting recurrence, and developing a conjecture in \cite{diaconis1988}, they also show that 
an ERRW can be represented as a random walk on a graph with random edge weights (a random walk on a random environment), 
whose distribution is the limit law of the fractions of time spent by the process on the edges. Recalling the representation of reversible Markov chains as random walks on undirected graphs (see e.g. \cite{diaconisRolles2006}), this result implies that ERRWs are mixtures of {\em reversible} Markov chains, and leads to the mixing distribution.  

To extend the results to infinite graphs, more refined techniques are needed. Merkl and Rolles \citep{merklRolles2007a} proved that, for any {\em locally finite} graph, an ERRW is a mixture of reversible  Markov chains, irrespectively whether it is recurrent or not. To show this, they first obtain more detailed  properties of the random environment for a finite graph, then treat the infinite case by approximating the infinite (but locally finite) graph with a sequence of finite graphs. 
In the finite case, they study the predictive distribution of successor states of a state $i$, as considered here in Theorem \ref{conv}. In this  theorem, we show that such sequence of predictive distributions converges to the random transition $\tP_{i,j}$, while \cite{merklRolles2007a}  give more refined results on the limit law. By comparing the predictive distribution of successor states with a Polya urn, whose asymptotic behaviour is well understood, they obtain bounds on the tails of the distribution of the random transition probabilities $\tP_{i,j}$. 
Such bounds are uniform in the size of the graph. Thus, they can be  exploited in the approximation of the infinite graph by finite graphs, to show that the ERRW is a mixture of reversible Markov chains, by also taking into account that, for a reversible Markov chain, there is a relationship between the transition probabilities and the edge weights in its graph representation.  
Although the work by Merkl and Rolles \citep{merklRolles2007a} does not deal with recurrence explicitly, their results give a valuable contribution to the study of recurrence for the {\em infinite} case. A  reversible irreducible Markov chain is recurrent if and only if, in its graph representation, the sum of the edge weights is finite. Thus, once it is proved that an ERRW can be represented as a random walk on a random environment, controlling the tails of the  distribution of the random edge weights is a technique to show that they are strictly positive and summable, almost surely and, therefore, the process is recurrent. Proving such bounds is technically hard. Extending results in \cite{merklRolles2005} and \cite{rolles2006}, Merkl and Rolles \citep{merklRolles2007b} study the asymptotic behavior of ERRWs on general multi-level ladders; in particular, they show that the edge weights decay exponentially in space, and prove recurrence.  Developing on the results in \cite{merklRolles2007a}, 
Merkl and Rolles \citep{merklRolles2009} prove recurrence of ERRW on a large class of periodic graphs satisfying certain symmetry properties. 
A different technique based on a representation of ERRWs in terms of vertex-reinforced jump process is used by Sabot and Tarres \citep{sabotTarres2014}, who show that ERRWs on $\mathbbm{Z}^d$ are strongly recurrent for any $d$, for large reinforcement, under mild conditions. 

Although reversible Markov chains have specific features, some of these techniques, possibly the approximation by finite graphs to obtain detailed results on the limit law of the predictive  distribution of successor states, could be helpful for other Markov exchangeable processes.

\section{Colored edge reinforced random walks}  \label{sec:examples}

In this section, we provide an illustration of the previous results through a
novel predictive construction. To some extent, the predictive scheme  proposed here is a generalization of ERRWs, with colored edges. Introducing colors  allows to reinforce groups of edges, even if they are not crossed. Colors' reinforcement could be exploited to express restrictions or global properties of the process. We use Theorem \ref{pred} to study the conditions under which this predictive scheme characterizes a Markov exchangeable process, and
provide examples which satisfy such conditions. These include processes found in the literature as well as novel characterizations.

Informally, the process is described as a random walk on a colored graph where edges and colors are reinforced when  crossed.
Consider a directed graph with vertices in a finite or countable set $S$ and with edge set $E$. Each directed edge $(i,j)$ in $E$ is given a
color $c(i,j)$ from a set of colors $C$ and a weight $\beta_{i,j}>0$.
 Each color $c$ is assigned an initial weight $\alpha_c>0$.
For each vertex $i$, let $C(i)$ denote the set of colors of all the edges starting at $i$ and let $\alpha_{C(i)}=
\sum_{c \in C(i)} \alpha_c$ be the overall weight of all colors from $i$.  For each color $c$, let $E_c$ denote the set of all edges of color $c$ in the graph, and
$\beta_{i,E_c}=\sum_{j: (i,j) \in E_c} \beta_{i,j}$ be the overall weight  of all edges of color $c$ starting from $i$.
We assume $\alpha_{C(i)}<\infty$ and $\beta_{i,E_c}<\infty$ for every $i$ and every $c$.

Let  $x_0$ be the starting point of the random walk. At step one, a color $c$ is selected among the colors in $C(x_0)$, with probability $\alpha_{c}/ \alpha_{C(x_0)}$. Then, a directed edge $(x_0, x_1)$ is selected among the edges of color $c$ from $x_0$, with probability
$\beta_{x_0,x_1}/ \beta_{x_0, E_{c}}$. The process moves to $X_1=x_1$ and both the weight of color $c$ and of edge $(x_0,x_1)$ are incremented by one. The walk is repeated in the same way, this time starting from $x_1$ and with the new weights; and so on for the following steps.

Let $\bX$ denote the process generated in this manner, and let $\TT(x_0,\bX_{1:n})$ be its transitions count process, with
$\TT_{i,j}$ denoting the number  of transitions across edge $(i,j)$ in $(x_0, \bX_{1:n})$. The  resulting predictive probabilities for the process $\bX$ are as follows. For $(i,y)\in
E_c$, 
\begin{equation}  \label{predwalk}
 p(y|x_0,{\bf x},i)= 
 \frac{\alpha_{c}+{\bf T}_{E_c}(x_0,{\bf x},i)}{\alpha_{C(i)}+{\bf T}_{C(i)}(x_0,{\bf x},i)} \;
\frac{\beta_{i,y}+{\bf T}_{i,y}(x_0,{\bf x},i)}{\beta_{i, E_c}+ {\bf T}_{i, E_c}(x_0,{\bf x},i)}
\end{equation}
where, for each color $c$, $\TT_{E_c}=\sum_{(u,v) \in E_c} \TT_{u,v}$ is the number of transitions over edges of color
$c$, and for each vertex $i$, $\TT_{C(i)}=\sum_{c \in C(i)} T_{E_c}$ and $\TT_{i, E_c}=\sum_{u: (i,u) \in E_c} T_{i,u}$. Notice that, if there is only one edge of color $c$ from $i$, the second factor in (\ref{predwalk}) is equal to one, thus the predictive probabilities reduce to the color updating. Analogously, if all edges from $i$ have the
same color, the first factor in (\ref{predwalk}) is one and the predictive probabilities reduce to the
edge's weight updating. Also, one could allow reinforcements different from one, or no reinforcement at all, and
this could be done both for colors and for edges.

\begin{prop} \label{prop:condGenerale}
The sequence of predictive rules (\ref{predwalk}) defines a Markov exchangeable process if and only if
\begin{equation} \label{eq:condGenerale}
\prod_{l=2}^m [\alpha_{C(y_{l-1})}+ \TT_{C(y_{l-1})}(x_0,{\bf x},i,y_1, \dots,y_{l-1})] = \prod_{l=2}^m
[\alpha_{C(y'_{l-1})}+ \TT_{C(y'_{l-1})}(x_0,{\bf x},i,y'_1, \dots,y'_{l-1})]
\end{equation}
for every $i$, $\bx$ and $\by=(y_1, \ldots, y_m)$,  $\by'=(y_1', \ldots,y_m') $ such that
$\by=(\bu, \bw, i, \bv, \bw,i)$ and $\by'=(\bv, \bw, i, \bu, \bw,i)$, with $\{\bu\}, \{\bv\}, \{\bw\}, \{i\}$
disjoint.
\end{prop}

\noindent {\sc Proof}. The sequence of predictive rules (\ref{predwalk}) defines a Markov exchangeable process
if and only if it satisfies conditions $a)$ and $b)$ of Theorem \ref{pred}. Condition $a)$ is immediate, as
(\ref{predwalk}) depends on $(x_0,\bx,i)$ only through the transitions and the last state $i$. Direct computation of $p(\by \mid x_0,\bx,i)$ involves the product of terms of the form (\ref{predwalk})
recursively updated, and one can easily check that condition $b)$ is satisfied if, for vectors $\by , \by'$ of
the form $\by=(\bu, \bw, i, \bv, \bw,i)$ and $\by'=(\bv, \bw,i, \bu, \bw,i)$ with $\{i\},\{\bu\}, \{\bv\},
\{\bw\}$ disjoint, the following equivalences hold:
\begin{equation} \label{eq:b1}
\frac{\prod_{l=1}^m [\alpha_{c(y_{l-1}, y_l)}+ \TT_{E_{c(y_{l-1}, y_l)}}(x_0, \ldots, y_{l-1})]}
{\prod_{l=1}^m [\alpha_{C(y_{l-1})}+ \TT_{C(y_{l-1})}(x_0,\dots,y_{l-1})]}
= \frac{\prod_{l=1}^m [\alpha_{c(y'_{l-1}, y'_l)}+ \TT_{E_{c(y'_{l-1}, y'_l)}}(x_0, \dots, y'_{l-1})]}
{\prod_{l=1}^m [\alpha_{C(y'_{l-1})}+ \TT_{C(y'_{l-1})}(x_0,\dots, y'_{l-1})]} ;
\end{equation}
\begin{eqnarray}  \label{eq:b2}
&&\frac{\prod_{l=1}^m [\beta_{y_{l-1},y_l} +\TT_{y_{l-1}, y_l}(x_0,\dots,y_{l-1})]} {\prod_{l=1}^m
[\beta_{y_{l-1}, E_{c(y_{l-1}, y_l)}}+
\TT_{y_{l-1}, E_{c(y_{l-1}, y_l)}}(x_0,\dots, y_{l-1})]} \nonumber \\
&=&
\frac{ \prod_{l=1}^m  [\beta_{y'_{l-1},y'_l}+\TT_{y'_{l-1}, y'_l}(x_0,,\dots,y'_{l-1})]}
{\prod_{l=1}^m [\beta_{y'_{l-1}, E_{c(y'_{l-1}, y'_l)}}+ \TT_{y'_{l-1}, E_{c(y'_{l-1},
y'_l)}}(x_0,\dots,y'_{l-1})]}
\end{eqnarray}
where $y_0=y_0'=i$. 
The factors on both sides of (\ref{eq:b2}) only depend on the last state and the transitions from it.
Similarly to the proof of Corollary \ref{tindip}, one can see that (\ref{eq:b2}) always holds. 
As for condition (\ref{eq:b1}), the numerators on the left hand side only depend on the number of times colors
$c(y_{l-1}, y_l)$ are visited in $(x_0,\bx,i, \by)$, which remain unchanged in $(x_0,\bx,i, \by')$. Thus,
condition $b)$ of Theorem \ref{pred} is satisfied if the denominators on both sides of (\ref{eq:b1}) are equal,
that is equivalent to (\ref{eq:condGenerale}). Since (\ref{eq:b2}) is always true, condition 
(\ref{eq:condGenerale}) is also necessary. \hfill $\diamond$

\medskip

Condition (\ref{eq:condGenerale}) has to be checked case by case, depending on the structure of the graph and of the reinforcement. 
Reinforced Hoppe's urns and ERRWs are special cases of the predictive scheme (\ref{predwalk}), for which (\ref{eq:condGenerale}) holds.

Reinforced Hoppe's urn schemes discussed in Example \ref{hoppe} are a particular  case of (\ref{predwalk}) for which all edges in the graph have the same color. Then, the predictive rule (\ref{predwalk}) reduces to 
(\ref{eq:Hoppepred}); for which condition  (\ref{eq:condGenerale}) is immediate. Some extensions will be given in Section \ref{sec:partitionColors}. 

ERRWs are defined for undirected graphs, but they can be framed in our scheme by assigning to every pair of directed edges $(i,j)$ and $(j,i)$ the same color, with a different color for each different pair, and augmenting the graph by associating to each vertex $i$ an auxiliary vertex, say $i^*$, to represent loops $(i,i)$ as the pair
of directed edges $(i, i^*), (i^*, i)$. Then, when edge $(i,j)$ is crossed, both $(i,j)$ and $(j,i)$ are
reinforced by one, and loops are reinforced by $2$. In this case, the predictive rule (\ref{predwalk}) reduces
to the predictive rule that characterizes ERRWs. Markov exchangeability can be easily verified through condition (\ref{eq:condGenerale}), as it only involves disjoint vectors $\bv, \bw, \bw, i$, and, in this case,  colors between the vertices of disjoint vectors are all distinct. Thus, for any element $u_l$ of $\bu$
$$ \TT_{C(u_l)}(x_0,\bx, i, u_1, \ldots, u_l)=
\TT_{C(u_l)}(x_0,\bx, i, \bv, \bw, i, u_1, \ldots, u_l), 
$$
and similarly for $\bv$. For vertex $i$, the terms involved
in (\ref{eq:condGenerale}) are $\TT_{C(i)}(x_0,\bx,i)$ and \\
$\TT_{C(i)}(x_0,\bx,i,\bu, \bw, i)=\TT_{C(i)}(x_0,\bx, i,
\bv, \bw, i)= \TT_{C(i)}(x_0,\bx, i)+2$. Similarly for $\bw$.

\subsection{Colored edge reinforced random walks with partitioned colors} 
\label{sec:partitionColors} 
Suppose that the graph's colors are partitioned into groups
$\{C_1, \ldots, C_N\}$, $N \leq \infty$, such that, for all $i$, there exist $m \leq N$ such that $C(i)=C_m$. In other words, for every  pair of vertices $i$ and $j$, either $C(i)=C(j)$ or $C(i)\cap C(j)=\emptyset$. 
As a consequence, the left hand side of  (\ref{eq:condGenerale}) only depends on the transitions through edges with 
colors in the sets $C_1, \ldots, C_N$ in $(i,\by)$, which are the same for $(i,\by')$ if
$\by=(\bu,\bw,i,\bv,\bw,i)$ and $ \by'=(\bv,\bw,i,\bu,\bw,i)$. Thus, (\ref{eq:condGenerale}) holds and the
process $\bX$ is Markov exchangeable.

The process is recurrent if 
$\inf_i \alpha_{c(i, x_0)} >0$ and $\inf_i \beta_{i,x_0} >0$, by Theorem \ref{th:rec}. Recurrence holds under a milder restriction if the state space is finite. Let $Q=[Q_{i,j}]$ be the matrix  of normalized edge weights $Q_{i,j}=\beta_{i,j}/\sum_{j'} \beta_{i,j'}$, with $Q_{i,j}=0$ if the edge $(i,j)$ is not present in the graph. 

\begin{prop}\label{prop:recurrence-gERRW}
Let $\bX$ be a colored edges reinforced random walk with partitioned set of colors, finite state space and irreducible weight matrix $Q$. Then $\bX$ is recurrent.
\end{prop}
The proof is given in the Appendix. One first shows that, in this case, if a set of colors $C_m$ is visited infinitely often, then every color in $C_m$ is visited infinitely often, almost surely; and if a set of edges $E_c$ is visited infinitely often, every edge in $E_c$ is visited infinitely often, almost surely. 

It follows from Proposition \ref{prop:recurrence-gERRW} that a colored edge reinforced random walk $\bX$ with partitioned set of colors, finite state space and irreducible weight matrix $Q$ is a mixture of recurrent Markov chains. The prior is the  limit probability law of the predictive distributions, as discussed in Section \ref{sec:pred}.  Consider a state $i$ with $C(i)=C_m$ for some $m \in \{1, \ldots, N\}$. 
From expression (\ref{predwalk}), in the notation of Theorem \ref{conv}, 
$P(X_{\tau_n(i)+1}=j \mid {\cal F}_{\tau_n(i)})$ is the product of two terms, the predictive probability of choosing the color, say $c$, of the edge $(i,j)$, times the predictive probability, given ${\cal F}_{\tau_n(i)}$ and the color $c$ chosen, of picking the edge $(i,j)$ among the edges of color $c$ from $i$. One can easily see that the two terms have the expression of the predictive probabilities of having next color $c$ and next state $j$ for two independent exchangeable sequences, respectively:  an exchangeable sequence of colors in $C_m$, with directing measure $(\tP_m(c), c \in C_m)$ having Dirichlet distribution with parameters  $(\alpha(c), c \in C_m)$, and an exchangeable sequence of states in $A_{i,c}=\{j' : c(i,j')=c\}$, with directing measure $(\tP(j \mid i,c), j \in A_{i,c})$ having Dirichlet distribution with parameters $(\beta_{i,j}, j \in A_{i,c})$, independently of $\tP_m$.  
Because, for exchangeable sequences, the predictive distributions converge almost surely to the directing measure, it follows that $P(X_{\tau_n(i)+1}=j \mid {\cal F}_{\tau_n(i)})$ converges almost surely to 
$$
\tP_{i,j}= \tP_m(c(i,j)) \; \tP(j \mid i,c(i,j)).
$$ 
Thus, for a state $i$ with $C(i)=C_m$, the $i$th row of the random transition matrix, as a measure on $S$,  is given by 
\begin{equation} \label{eq:prior}
\tilde{P}_{i}(\cdot)=\sum_{c \in C_m} \tilde{P}_m(c) \, \sum_{y\in A_{i,c}} \tilde{P}(y
\mid i,c) \, \delta_y(\cdot), 
\end{equation} 
where
$(\tP_m(c), c \in C_m) \sim \, \mbox{Dirichlet}(\alpha(c), c \in C_m)$, 
$(\tP(j \mid i,c), j \in A_{i,c}) \sim \, \mbox{Dirichlet}
(\beta_{i,j}, j \in A_{i,c})$ for any $c \in C_m$, and  
$\tP_m$ and the $\tP(\cdot \mid i,c)$ are independent. 
In other words, the colors of the edges from $i$ determine a partition of the sample space, and the transition probability $\tP_i(\cdot)$ from $i$ first picks a region $A_{i,c}$ in the partition, with probability $\tP_m(c)$, then selects a state $y \in A_{i,c}$, with probability  $\tilde{P}(y \mid i,c)$. 
The rows of the random transition matrix are in general dependent, through the common component $\tilde{P}_m(c)$. 
The same results hold for infinite state space if the process $\bX$ is recurrent, with the understanding that, if $C_m$ or $A_{i,c}$ are countable, the Dirichlet distributions above become the appropriate Dirichlet processes. 

Regarding the transition  probability law $\tilde{P}_{i}$ as a probability measure on $C(i) \times S$, the
prior reveals some analogies with the Enriched Dirichlet process \citep{wade2011} for a bivariate random distribution, which
arises as a nonparametric extension of the generalized Dirichlet distribution.  

Let $\mu(\cdot ; x_0, \balpha, \bbeta)$ be the prior distribution defined by (\ref{eq:prior}) for a colored ERRW with weights $\balpha, \bbeta$, and let $\cal D$ denote the family of  such prior distributions for different starting values and weights. The family $\cal{D}$ is closed under sampling, that is, the posterior distribution still belongs to $\cal{D}$. Moreover, there exists a closed-form mapping to the posterior parameters (sometimes called {\em functional conjugacy}; see the discussion in \cite{orbanz2009}): 
if $\bX$ has prior $\mu(\cdot ; x_0, \balpha, \bbeta)$, then the posterior distribution, given $\bX_{1:n}=\bx_{1:n}$, is
$\mu(\cdot; x_n,\balpha(x_0, \bx_{1:n}), \bbeta(x_0, \bx_{1:n}))$, where
\begin{equation} \label{eq:posterior}
\balpha(x_0, \bx_{1:n})=\alpha_c+\TT_{E_c}(x_0,\bx_{1:n}), 
\quad 
\bbeta(x_0, \bx_{1:n})_{i,j}=\beta_{i,j}+\TT_{i,j}(x_0,\bx_{1:n}) .
\end{equation}
This implies that the posterior distribution belongs to $\cal D$, thus $\cal D$ is closed under sampling. To prove (\ref{eq:posterior}), notice that the conditional distribution of $(X_{n+1}, X_{n+2}, \ldots)$, given $\bX_{1:n}=\bx_{1:n}$, is the distribution of the same colored edge reinforced random walk, but with initial state $x_n$ and weights $\balpha(x_0, \bx_{1:n}), \bbeta(x_0, \bx_{1:n})$.

The following examples provide further insights into the nature of the process.

\begin{exa} Independent enriched Dirichlet rows.
{\em Suppose that the set of colors associated to distinct vertices are all different, i.e. $C(i) \cap
C(j)=\emptyset$ for $i \neq j$.
In this case, no probabilistic dependence is induced through the predictive distributions and the resulting
random transition matrix has  independent rows. As the Dirichlet process, the prior distribution on the rows is closed under
sampling, but allows more flexibility in having the choice of two scale parameters, rather than just one. For
example, suppose that the graph represents a physical network whose nodes are partitioned in local nets $A_1,
\ldots, A_k$, with $\bX$ describing some flow of information through the network, and assume one wishes 
to express the prior information that, from a node $i$, many local networks $A_{m}$ are visited, but only a few
states inside each local network tend to be visited. Such prior information could not be expressed by a Dirichlet
process, but, by the clustering properties of Dirichlet processes, it could be incorporated in the prior
(\ref{eq:prior}) by choosing a large value of the normalizing constant  $\alpha_{C(i)}$ and  small values for $\beta_{i,E_c}$.}
\end{exa}

\begin{exa} \label{ex:priorConstraints} Analytic constraints. 
{\em The case where some vertices share the same group of colors is somehow opposite  to the previous example. The predictive probabilities (\ref{predwalk}) imply analytic constraints on the random transition matrix. Indeed, from (\ref{eq:prior}), we obtain
	\begin{equation} \label{eq:constrains}
\sum_{y \in A_{i,c}} \tilde{P}_{i,y}= \sum_{y \in A_{i',c}} \tilde{P}_{i',y} = \tilde{P}_m(c) .
\end{equation}
Thus, sums, by row, of probabilities of transitions along edges of the same color are constant. As a simple example, consider a graph with loops. Let the loops be colored in red and all other edges be colored in
blue. The predictive distributions induce a mixture of random walks with equal transition probabilities on
loops.}
\end{exa}

\begin{exa} \label{ex:dummy} Colored edge reinforced random walks with dummy states. 
{\em One way of introducing dependence among the rows of the random transition matrix, without strictly imposing constraints as those of expression  (\ref{eq:constrains}), is to augment the graph $G$ with auxiliary dummy states and edges, as follows. 
Consider a graph $G$, that can have partitioned colors or not. For an edge $(i,j)$ in $G$, let us introduce a new vertex $ {i}^*$ between $i$ and $j$, by adding a state $i^*$ together with the edges $(i, {i}^*)$ and $( {i}^*,j)$. The construction is repeated, adding one or more dummy vertices for each edge $(i,j)$ in a set $I$.  Let $I^*_{i,j}$ denote the set of dummy states between  $i$ and $j$; $I^*$ the set of dummy states and  $S^*=S\cup I^*$. 

If the  augmented graph $G^*$ has partitioned colors, then an ERRW $\bX^*$ on $G^*$, with weights $\balpha^*, \bbeta^*$, and starting at $x_0\in S$, is Markov exchangeable. Moreover, if recurrent, $\bX^*$ is a mixture of Markov chains with prior distribution $\mu^*(\cdot; x_0, \balpha^*, \bbeta^*)$ on the random transition matrix $\tP^*$.

Now, let $\bX$ be the process defined by deleting the dummy states from the paths of $\bX^*$. The process $\bX$ is well defined and its probability distribution can be recovered from the law of $\bX^*$, as
\begin{equation}
\label{eq:dummy_sum}
p(\bx)=P(\bX_{1:n}=\bx)=\sum_{(x_0,\bx^*)\in A(x_0,\bx)} P( {\bX}^*_{1:n+m}= {\bx}^*)= \sum_{(x_0,\bx^*)\in A(x_0,\bx)} p^*(\bx^*), 
\end{equation}     
 where $(n+m)$ is the length of $\bx^*$ and the sum is taken over the set $A(x_0,\bx)$ of all sequences $(x_0,{\bx}^*)$ consistent with $(x_0,\bx)$; that is, sequences that start and end as $(x_0,\bx)$ and lead to  $(x_0,\bx)$ when deleting the dummy states. The process $\bX$ is Markov exchangeable. If $(x_0,\bx) \sim (x_0, \bx')$, it is fairly simple to see that the elements of $A(x_0, \bx)$ and $A(x_0, \bx')$ are pairwise equivalent. Thus, $p(\bx)= p(\bx')$. 
 
Moreover, if $\bX^*$ is recurrent, then $\bX$ is recurrent, as well, because $x_0 \in S$. In this case, $\bX$ is a mixture of recurrent Markov chains, and the random transition probabilities $\tP_{i,j}$ can be obtained by the transformation  
$$ 
\tP_{i,j} = \tP^*_{i,j} + \sum_{i^* \in I^*_{i,j}} \tP^*_{i,i^*} 
$$
of the random transition probabilities $[\tP^*_{i,j}] \sim \mu^*(\cdot; x_0, {\mathbf{\alpha}}, \mathbf{\beta})$ for the process $\bX^*$. 
The prior distribution for $\tP$ redistributes the masses assigned by $\mu^*$ according to the dummy states that have been introduced, and 
generally expresses correlation across the rows of the transition matrix. 

\medskip

As an illustration, consider a monochromatic finite graph $G$, of color $c_1$, with $k$ vertices, including loops. As discussed before, a colored ERRW on $G$ with edge weights $\beta_{i,j}$ is a Hoppe reinforced urn process with $\alpha_i q_i(j)=\beta_{i,j}$. 
When recurrent, such process is a mixture of Markov chains, with independent Dirichlet($\beta_{i,1}, \ldots, \beta_{i,k}$) distributions on the rows of the transition matrix.  
Let us augment the graph by adding, for every vertex $i$, a dummy vertex $i^*$ along with edges  $(i,i^*)$ and $(i^*,i)$. We color all  edges $(i,i^*)$ with a single color $c_2\neq c_1$, and  the edges $(i^*,i)$ with a color $c_3$ different from $c_1$ and $c_2$. This results in partitioned colors on the extended graph $G^*$, the partition being $(C_1=(c_1, c_2), C_2=(c_3))$. Let $\bX^*$ be a recurrent colored ERRW on the augmented graph, and $\bX$ defined from $\bX^*$ as discussed above. Then, $\bX$ is a mixture of Markov chains with random transition probabilities:  
\begin{eqnarray*}
\tP_{i,j} &=& \tP^*_{i,j} \; = \tP_1^*(c_1) \, \tP^*(j \mid i,c_1) 
\quad  \quad  \mbox{for $j \neq i$}, \\
\tP_{i,i} &=& \tP^*_{i,i} + \tP^*_{i,i^*} = 
\tP_1^*(c_1) \, \tP^*(i \mid i,c_1) + (1-\tP_1^*(c_1))  
\end{eqnarray*}
The $i$th row of the transition matrix $\tP$ is a mixture of the probability distribution $(\tP^*(j \mid i, c_1), j \in S) \sim$ 
Dirichlet($\beta^*_{i,1}, \ldots, \beta^*_{i,k}$), the same distribution as for a Hoppe reinforced urn process, and a degenerate distribution on $i$, with mixing weight $\tP^*_1(c_1) \sim$ Beta($\alpha^*_{c_1}, \alpha^*_{c_2}$), independent of the $\tP^*(j \mid i,c_1)$.
The probability of a loop is a weighted average of $\tP^*(i \mid i, c_1)$ and $\tP^*(i^* \mid i, c_2)=1$, thus it is higher than for the corresponding reinforced urn process. The rows of the transition matrix are correlated,  due to the common random variable $\tP_1^*(c_1)$, but no strict equality is imposed, differently from Example \ref{ex:priorConstraints}. 

\medskip

As a further example, dummy states can be used to express prior  inequalities between the elements of the random transition matrix. Suppose that, in the original graph $G$,  edges $(i,j)$ and $(i',j')$,  $j$ and $j'$  distinct or not, have the same color $c_1$ and are the only edges from $i$ and from $i'$ with color $c_1$. A dummy state $i^*$ is added  between $i$ and $j$, together with edges $(i,i^*)$ and $ (i^*,j)$. The edge $(i,i^*)$ is given a color $c_2\neq c_1$. If necessary, other dummy vertices are added, until the extended graph has partitioned colors, never including edges of color $c_1$ from $i$ or from $i'$, nor dummy vertices between $i'$ and $j'$. The processes $\bX^*$ and $\bX$ are constructed as above. The random transition probabilities for the process $\bX$ satisfy
	$$
	\tP_{i,j}=\tP^*_1(c_1) +\tP^*_1(c_2) \tP^*(j \mid i,c_2); \quad \tP_{i',j'}=\tP^*_1(c_1).
	$$
Hence, $\tP_{i,j} > \tP_{i',j'}$. 

\medskip

The construction by dummy states can be quite flexible in expressing prior beliefs. Notice that it is always possible to augment a graph $G$ until the extended graph has partitioned colors, hence, the proposed construction can be used to define a Markov exchangeable reinforced random walk on any graph $G$. However, while the prior distribution $\mu^*$ for the process $\bX^*$ is closed under sampling, this is no longer true for the prior $\mu$ for $\bX$. In principle, computing the posterior distribution $\mu(\cdot \mid \bx)$ on the random transition matrix $\tP$ for $\bX$, given observations $\bx=\bx_{1:n}$, remains simple,  since  
\begin{equation}  \label{eq:posterior*}
\mu(\;\cdot \mid \bx)= \sum_{(x_0,\bx^*) \in A(x_0,\bx)} 
\mu^*\left(\;\;\cdot\;\;; \, x_n, \balpha^*(x_0, \bx^*), \bbeta^*(x_0,\bx^*)\right)\, 
p_{X^* \mid X}(\bx^* \mid \bx), 
\end{equation}
where the conditional probability $p_{X^* \mid X}(\bx^* \mid \bx)= 
p^*(\bx^*)/p(\bx)$ can be easily computed from $p^*$ and using  (\ref{eq:dummy_sum}). The summation in (\ref{eq:dummy_sum}) can  actually be rearranged into a smaller number of terms. The set $A(x_0, \bx)$ can be partitioned in classes of equivalent vectors $A_\bmm=\{ (x_0, \bx^*) \in A(x_0, \bx^*) : \bmm(x_0, \bx^*)= \bmm \}$, 
where 
$\bmm(x_0, \bx^*)=\bmm(x_0, \bx^*)=(m_{i^*}(x_0, \bx^*), i^* \in I^*)$ and  $m_{i^*}(x_0, \bx^*)$ denotes the number of visits to the dummy state $i^*$ in $(x_0, \bx^*)$. For any $\bmm$, all the strings in 
$A_\bmm$ are equivalent and, therefore, have the same probability. 
A simple combinatorial argument shows that the cardinality of $A_\bmm$ is 
$$
N_\bmm = \prod_{(i,j)\in I}\frac{\TT_{i,j}(x_0,\bx)!}
{(\TT_{i,j}(x_0,\bx)-m^*_{i,j})! \prod_{i^*\in I^*_{i,j}} m_{i^*}!},
$$
where $m^*_{i,j}=\sum_{i^*\in I^*_{i,j}}m_{i^*}$. 
Thus, 
  \begin{equation}  \label{eq;dummyprob}
  p(\bx)=\sum_{\bmm} 
  p^*(\bx^*_{\bmm}) \,  N_{\bmm}, 
  \end{equation} 
where $(x_0,x^*_{\bmm})$ is any string in $A_\bmm$. Although reduced with respect to (\ref{eq:dummy_sum}), the number of terms in (\ref{eq;dummyprob})  still explodes for long paths,  or when the number of dummy states is large. Yet, simulation techniques can be used to sample from 
$p_{X^*\mid X}(\bx^* \mid \bx)$ and to compute a Monte Carlo approximation of (\ref{eq:posterior*}). To simulate from $p_{X^*\mid X}$, one has to simulate the possibly missing passages through dummy states. Notice that a string $(x_0, \bx^*)$ in $A(x_0, \bx)$ can be described in terms of the successors states: 
each state $x_i \notin I$ in the observed sample $(x_0,\bx)$ has a known successor state in $\bx^*$; 
while for each observed pair of consecutive values $(x_i, j)$, where $x_i \in I$, the successor of $x_i$ in $\bx^*$ can be any state in the set $\{ j, I^*_{x_i,j}\}$. By construction, the successor of a dummy state $i^* \in I^*_{i,j}$ is always equal to $j$. 
Thus, generating $\bx^*$ from $p_{X^* \mid X}(\bx^* \mid \bx)$
 is equivalent to generating the sequence of unknown successor states $\bV^*$ from the conditional distribution 
$P( \bV^* = \bv^* \mid \bV=\bv, \bV^* \in B)$, given the known successors  $\bV$ and the appropriate constraints on the remaining ones, denoted by $\bV^* \in B$. The simplest Gibbs sampling scheme consists in generating the $\bV^*$ one at the time from their full conditional distributions. Assume that we want to generate a successor $V_i$ of state $i$, corresponding to consecutive observations $(i,j)$ in $(x_0,\bx)$. By partial exchangeability of the matrix of successor states, one can permute the successors so that $V_i$ is the last successor of state $i$. Then, its full conditional distribution selects a state $k$ in $\{j, I^*_{i,j}\}$  with probability proportional to 
 $$ 
 \frac{\alpha^*_{c(i,k)} + \TT_{c(i,k)}} {\alpha^*_{C(i)} + \TT_{C(i)}}
 \; 
 \frac{\beta^*_{i,k} + \TT_{i,k}}
 {\beta^*_{i,E_{c(i,k)}} + \TT_{i,E_{c(i,k)}}}
$$ 
where the transitions counts are computed on all the other successor states.  
}
\end{exa} 
 
  \appendix
 
 \section{Complements and proofs} \label{appendix}
 
 \subsection{Complements for Section \ref{sec:preliminary}}
 
 We following lemma is used for the proof of Theorem \ref{conv}.
 For a state $i$, let ${\cal F}_{V_{i,n}^\checkmark}$ denote the sigma algebra generated by $(V_{i,k},k<n;V_{j,l},j\in S^*,j\neq
 i,l\geq 1)$. 
 \begin{lemma} \label{check}
 	Under the hypotheses of Theorem \ref{conv},  ${\cal F}_{\tau_n(i)} \subseteq {\cal F}_{V_{i,n}^\checkmark}$. 
 \end{lemma}
 
 \noindent {\sc Proof.} Let us consider $B\in {\cal F}_{\tau_n(i)}$ and  show that $B\in {\cal F}_{V_{i,n}^\checkmark}$. One can express $B$ as 
 $$
 B=\cup_{k=1}^\infty ( B \cap (\tau_n(i)= k) \cup (B \cap (\tau_n(i)=\infty) ).$$
 We show that the events: i) $B \cap (\tau_n(i)= k)$, $k\geq 1$ and  ii) $B\cap (\tau_n(i)=\infty)$ belong to
 ${\cal F}_{V_{i,n}^\checkmark}$.
 \\
 i) Let ${\cal F}_k$ be the sigma-field generated by $\bX_{1:k}$. Since $B \cap (\tau_n(i)= k)\in {\cal F}_k$, there exists $B_{n,k}$ such that 
 $$ 
 B\cap (\tau_n(i)\leq k)=\cup_{\bx_{1:k} \in B_{n,k}}(\bX_{1:k}=\bx_{1:k}).
 $$ 
 It is proved in \cite{fortini2002} that, for any $\bx_{1:k}$, there
 exist $m$, $s_1,\dots,s_m$, $n_{s_1},\dots,n_{s_m}$ and $v_{s_1,1},\dots,v_{s_1,n_{s_1}}, \dots, v_{s_m,1},\dots,v_{s_m,n_{s_m}}$ such that
 \begin{eqnarray} \label{eq:xasv}
 	(\bX_{1:k}=\bx_{1:k}) = (V_{s_1,1}=v_{s_1,1},\dots,V_{s_1,n_{s_1}}=v_{s_1,n_{s_1}},\dots,V_{s_m,1}
 	=v_{s_m,1},\dots,V_{s_m,n_{s_m}}=v_{s_m,n_{s_m}}).
 \end{eqnarray}
 For $\bx_{1:k} \in B_{n,k}$, one has $(\bX_{1:k}=\bx_{1:k}) \subseteq (\tau_n(i)= k)$, thus $n_i=n-1$; therefore, $B\cap
 (\tau_n(i))= k) \in {\cal F}_{V_{i,n}^\checkmark}$.  
 \\
 ii) To prove that $B\cap (\tau_n(i)=\infty)\in {\cal F}_{V_{i,n}^\checkmark}$, let 
 ${\cal C}=\{A \in{\cal
 	F}: A \cap (\tau_n(i)=\infty) \in {\cal F}_{V_{i,n}^\checkmark}\}.
 $
 It is easy to verify that $\cal C$ is a
 sigma-algebra. Furthermore, for every $k$ and $\bx_{1:k} \in S^k$, the event $(\bX_{1:k}=\bx_{1:k})$ belongs 
 to ${\cal C}$. Indeed, if $m$, $s_1,\dots,s_m$, $n_{s_1},\dots,n_{s_m}$ and $v_{s_1,1},\dots,v_{s_1,n_{s_1}}, \dots, v_{s_m,1},\dots,v_{s_m,n_{s_m}}$ satisfy (\ref{eq:xasv}), then 
 $(\bX_{1:k}=\bx_{1:k}) \cap (\tau_n(i)=\infty) = \emptyset$ 
 if $n_i\geq n$ or if $n_i=n-1$ and $x_k=i$; while, for $n_i<n-1$ or for $n_i=n-1$, with $x_k\neq i$, one has 
 \[
 \begin{aligned}
 (\bX_{1:k}=\bx_{1:k})&\cap (\tau_n(i)=\infty)\\
 &=\left(\bigcap_{j=1}^m\bigcap_{l=1}^{n_{s_j}}(V_{s_j,l}=v_{s_j,l})\right)\cap \left(\bigcap_{s_j\neq i}\bigcap_{l=n_{s_j}+1}^\infty
 (V_{s_j,l}\neq i)\right)\cap \left( \bigcap_{s\neq s_1,\dots,s_m}\bigcap_{l=1}^\infty (V_{s,l}\neq i)\right).
 \end{aligned}
 \]
 Therefore, for every $k$ and  $\bx_{1:k}$, the event $(\bX_{1:k}=\bx_{1:k})$ belongs to ${\cal C}$. It follows that ${\cal F}_k\subseteq {\cal C}$ for every $k$. Since $\cal C$ is a sigma-algebra and includes
 $\cup_{k=1}^\infty {\cal F}_k$, it includes the sigma-algebra $\vee_k {\cal F}_k$ generated by $\cup_{k=1}^\infty {\cal F}_k$. Let $B\in {\cal F}_{\tau_n(i)}$. Since ${\cal F}_{\tau_n(i)}\subseteq \vee_k {\cal F}_k$, then $B\in {\cal C}$; therefore, $B\cap(\tau_n(i)=\infty)\in {\cal F}_{V_{i,n}^\checkmark}$. \hfill $\diamond$
 
 \medskip
 
 The following Corollary rephrases and proves Corollary \ref{cor_indep}.
 
 \begin{cor} \label{cor_indep2}
 	Let $\bX$ be a mixture of recurrent Markov chains with random transition matrix $\tP$. The rows of $\tP$  are
 	stochastically independent if and only if
 	$$
 	P(X_{\tau_n(i)+1}=j \mid \TT(x_0,\bX_{1:\tau_n(i)}))=
 	P(X_{\tau_n(i)+1}=j \mid \TT_i(x_0,\bX_{1:\tau_n(i)})) \quad
 	a.s..
 	$$
 	Furthermore, in this case,
 	$$
 	P(X_{\tau_n(i)+1}=j \mid \TT_i(x_0,\bX_{1:\tau_n(i)})=\mathbf t_i)=
 	P(V_{i,n}=j \mid \sum_{l=1}^{n-1}\delta_{V_{i,l}}(k)=\mathbf t_{i,k}:k\in S) 
 		$$
 	almost surely with
 	respect to the probability distribution of $\TT_i(x_0,\bX_{1:\tau_n(i)})$. 
 \end{cor}
 
 \noindent {\sc Proof.} 
 Suppose first that the rows of $\tP$  are stochastically independent.
 By Lemma  \ref{check},  
 ${\cal F}_{\tau_n(i)}\subseteq {\cal F}_{V_{i,n}^\checkmark}$ for $i \in S$, and by the independence assumption,
 $$
 P(V_{i,n}=j|{\cal F}_{V_{i,n}^\checkmark})=P(V_{i,n}=j|V_{i,1},\dots,V_{i,n-1}).
 $$
 Hence, the predictive probabilities from $i$ satisfy
 \begin{eqnarray} \label{eq:XtoV}
 	P(X_{\tau_n(i)+1}=j\mid {\cal F}_{\tau_n(i)})
 	&=&
 	E(P(V_{i,n}=j\mid {\cal F}_{V_{i,n}^\checkmark})\mid {\cal F}_{\tau_n(i)})\nonumber \\
 	&=& E(P(V_{i,n}=j \mid V_{i,1},\dots,V_{i,n-1}) \mid {\cal F}_{\tau_n(i)})\nonumber \\
 	&=&P(V_{i,n}=j \mid V_{i,1},\dots,V_{i,n-1}).
 \end{eqnarray}
 This proves that the transition probabilities from $i$ depend on
 $\bX_{1:\tau_n(i)}$ only through the transition counts from $i$. 
 
 Conversely,  suppose  that, for every $j$, $n$ and $\mathbf x_{1:n}$, the predictive probability
 $p(j \mid x_0, \bx_{1:n})$ is a function 
 $\pi(j \mid x_n, \TT_{x_n})$ of $x_n$ and of the transition counts $\TT_{x_n}$ from $x_n$. 
  Then $P(X_{\tau_n(i)+1}=j \mid {\cal F}_{\tau_n(i)})$ is a function of $V_{i,1},\dots,V_{i,n-1}$. 
 This implies that the rows
 of the array $(V_{i,n},i\in S,n\geq 1)$ are stochastically independent. Since the random transition probability $\tP_{i,j}$ is the limit of the relative frequency of $j$ in $V_{i,1},V_{i,2},\dots$, then the rows of 
 $\tP$ are stochastically independent as well.
 
 To prove the last assertion, notice that, if the sequences $(V_{i,n},n\geq 1)_{i\in S^*}$ are independent, then (\ref{eq:XtoV}) follows, and by exchangeability of $(V_{i,n},n\geq 1)$ we have the thesis.  
 \hfill $\diamond$

 \subsection{Complements for Section 3}

 \noindent {\sc Proof of Lemma \ref{lem3}}. The existence of a minimal element can be proved as for Lemma \ref{ii}. We want to prove
 that the minimal element $\bx^*=(\bu^*,k,j,\bv^*,k',j')$ satisfies:
 
 $1)$ $\{\bu^*,k\}\cap\{j,\bv^*,k'\}$ contains only one element, say $i$, that appears once in $(\bu^*,k)$, as
 the first element, and once in $(j,\bv^*,k')$;
 
 $2)$ if $i\neq j$, then $(\bv^*,k')$ contains no $j$ after $i$, and the strings before and after $i$
 in $\bv^*$ have no common elements.
 
 Let us prove $1)$. Had $(\bu^*,k)$ and $(j,\bv^*,k')$ more than one common element, we could delete the first
 part of $\bx^*$ and obtain a string in $\cal C$ that is shorter than $\bx^*$. Let us denote by $i$ the only common element.
 For the same reason, the state $i$ has to appear once in $(\bu^*,k)$, as its first element. For showing that $i$ appears
 only once in $(j, \bv^*, k')$, assume by contradiction that $i$ appears at least twice. Let us distinguish the following
 cases: $i=j$, $i=k$, $i=k'$ and $i\neq j,k,k'$. If $i=j$, then, by the above result, $j$ is the first element of
 $\bu^*$, say $\bu^*=(j, \bu^*_1)$; then $\bv^*$ cannot contain $j$, otherwise we could write
 $$\bx^*=(j,\bu_1^*,k,j,\bv_1^*,j,\bv_2^*,k',j)\sim (j,\bv_1^*,j,\bu_1^*,k,j,\bv_2^*,k',j)$$  and obtain a
 shorter string in $\cal C$ by deleting $(j,\bv_1^*)$. Similarly, for $i=k$, we could write $\bx^*=(k,\bu^*,k,j,\bv_1^*,k,\bv_2^*,k,\bv_3^*,k',j)\sim (k,\bv_2^*,k,\bu^*,k,j,\bv_1^*,k,\bv_3,k',j)$ and
 obtain a shorter element of $\cal C$ by deleting $(k,\bv_2^*)$. For $i=k'$, we could set $\bx^*=(k',\bu^*,k,j,\bv_1^*,k',\bv_2^*,k',j)\sim (k',\bv_2^*,k',\bu^*,k,j,\bv_1^*,k',j)$ and a shorter element
 could be obtained by deleting $(k',\bv_2^*)$.
 For $i\neq j,k,k'$, we could write 
 $\bx^*=(i,\bu^*,k,j,\bv_1^*,i,\bv_2^*,i,\bv_3,k',j)\sim (i,\bv_2^*,i,\bu^*,k,j,\bv_1^*,i,\bv_3,k',j)$ and we
 could delete $(i,\bv_2^*)$.
 
 Let us prove assertion $2)$. We distinguish the following cases: $i=k$; $i=k'$, and $i\neq k,k'$. Suppose,
 by contradiction, that $2)$ does not hold. Then:\\
 If $i=k$, we could write $\bx^*=(k,j,\bv_1^*,k,\bv_2^*,j,\bv_3^*,k',j)\sim (k,\bv_2^*,j,
 \bv_1^*,k,j,\bv_3,k',j)$ and obtain an element of $\cal C$ that is shorter than $\bx^*$ by deleting
 $(k,\bv_2^*)$; furthermore, we could write $\bx^*=(k,j,\bv_1^*,w,\bv_2^*,k,\bv_3^*,w,\bv_4^*,k',j)\sim
 (k,\bv_3^*,w,\bv_2^*,k,j,\bv_1^*,w,\bv_4^*,k',j)$ and  obtain a shorter element  by deleting $(k,\bv_3^*)$; \\
 If $i=k'$, we could write $\bx^*=(k',\bu^*,k,j,\bv_1^*,k',\bv_2^*,j,\bv_3^*,k',j)\sim
 (k',\bv_2^*,j,\bv_1^*,k',\bu^*,k,j,\bv_3^*,k',j)$ and 
 $\bx^*=(k',\bu^*,k,j,\bv_1^*,w,\bv_2^*,k',\bv_3^*,w,\bv_4^*,k',j)\sim
 (k',\bv_3^*,w,\bv_2^*,k',\bu^*,k,j,\bv_1^*,w,\bv_4^*,k',j)$, and 
 obtain shorter elements in $\cal C$ by deleting
 $(k',\bv_2^*)$ and $(k',\bv_3^*)$, respectively; \\
 If $i\neq k,k'$, we could write $\bx^*=(i,\bu^*,k,j,\bv_1^*,i,\bv_2^*,j,\bv_3,k',j)\sim (i,\bv_2^*,j,\bv_1^*,
 i,\bu^*,k,j,\bv_3^*,k',j)$ and $\bx^*=(i,\bu^*,k,j,\bv_1^*,w,\bv_2^*,i,\bv_3^*,w,\bv_4^*,k',j)\sim
 (i,\bv_3^*,w,\bv_2^*,i,\bu^*,k,j,\bv_1^*,w,\bv_4^*,k',j)$ and obtain shorter elements by deleting $(i,\bv_2^*)$
 and $(i,\bv_3^*)$, respectively. \hfill$\diamond$

 \medskip
 
 \subsection{Complements for Section \ref{sec:examples}}
 {\sc Proof of Proposition \ref{prop:recurrence-gERRW}}.
 Let $\bX$ be a colored edge reinforced random walk with partitioned colors $\{C_1,\dots,C_N\}$, finite state space and irreducible normalized weights matrix $Q$. We first show that
 
 (a) If a set of colors $C_m$ is visited infinitely often, then every color $c\in C_m$  is visited
 infinitely often, almost surely.
 
 (b) If a set of edges $E_c$ is visited infinitely often, then every edge in $E_c$ is
 visited infinitely often, almost surely. 
 \\
 Let us prove (a). For every $c\in C(i)$, let $A_{i,c}=\{j:c(i,j)=c\}$. We have
 \[
 P(X_{n+1}\in A_{i,c}|\bX_{1:n}=(\bx,i))=\frac{\alpha_c+{\bf T}_{E_c}(x_0,{\bf x},i)}{\alpha_{C(i)}+{\bf
 		T}_{C(i)}(x_0,{\bf x},i)}.
 \]
 Hence, if $c \in C_m$ and $\tau_n(C_m)=\inf\{t\geq \tau_{n-1}(C_m):C(X_t)=C_m\}$ is the time of the $n$th visit to a state with color set equal to $C_m$, we have 
 $$
 P(c(X_{\tau_n(C_m)},X_{\tau_n(C_m)+1})=c \mid \tau_n(C_m)<\infty)\geq \frac{\alpha_c}{\alpha_{C_m}+n}.
 $$
 It follows that
 \[
 \sum_{n=1}^\infty[ P(c(X_{\tau_n(C_m)},X_{\tau_n(C_m)+1})=c) \mid {\cal
 	F}_{\tau_n(C_m)}]\mathbbm{1}_{\{\tau_n(C_m)<\infty\}}+\mathbbm{1}_{\{\tau_n(C_m)=\infty\}}]=\infty \quad a.s.
 \]
 Reasoning as in the proof of Theorem \ref{rec2}, we obtain that, if $C_m$ is visited infinitely often, then every  $c\in C_m$ is visited infinitely often, almost surely. 
 \medskip
 \\
 (b). Let $(i,y)\in E_c$ and $\TT(x_0,\bx)=\TT$. Then
 \begin{eqnarray*}
 	& &P(X_{n+1}=y\mid \bX_{1:n-1}=\bx,c(X_n,X_{n+1})=c)\\
 	& &=\frac{\sum_{i:c(i,y)=c}P(X_{n+1}=y\mid\bX_{1:n}=(\bx,i))P(X_n=i\mid\bX_{1:n-1}=\bx)}
 	{\sum_{i:c(i,y)=c}P(c(X_n,X_{n+1})=c\mid \bX_{1:n}=(\bx,i))P(X_n=i\mid \bX_{1:n-1}=\bx)}\\
 	& &\geq \frac{\min_{i:(i,y)\in E_c}P(X_{n+1}=y\mid \bX_{1:n}=(\bx,i))}{\max_{i:(i,y)\in
 			E_c}P(c(X_n,X_{n+1})=c\mid \bX_{1:n}=(\bx,i))}\\
 	& &\geq \frac{\alpha_c+\TT_{E_c}}{\alpha_c+\TT_{E_c}+1} \; \frac{\min_{i:(i,y)\in E_c}\beta_{i,y}}{\max_{i:(i,y)\in
 			E_c}\beta_{i,y}+\TT_{E_c}}.
 \end{eqnarray*}
 Let $\tau_n(c)$ be the $n$th time the chain visits $E_c$. Then, for every $y$ such that $c(i,y)=c$ for some
 $i$,
 $$
 P(X_{\tau_n(c)+1}=y\mid \tau_n(c)<\infty) \geq \frac{\alpha_c+\TT_{E_c}}{\alpha_c+\TT_{E_c}+1}\frac{\min_{i:(i,y)\in
 		E_c}\beta_{i,y}}{\max_{i:(i,y)\in E_c}\beta_{i,y}+n}.
 $$
 Hence, 
 \[
 \sum_{n=1}^\infty[ P(X_{\tau_n(c)+1}=y|{\cal
 	F}_{\tau_n(c)})\mathbbm{1}_{\{\tau_n(c)<\infty\}}+\mathbbm{1}_{\{\tau_n(c)=\infty\}}]=\infty\quad a.s. .
 \]
 Again as in the proof of Theorem \ref{rec2}, we obtain that, if $E_c$ is visited infinitely often, then every edge in $E_c$ is visited infinitely often, almost surely. 
 
 To conclude the proof, let $i$ and $u$ be such that $Q_{i,u}>0$. If $i$ is visited infinitely often, then
 $C(i)$ is visited infinitely often; by a), every $E_c$ with $c\in C(i) $ is visited infinitely often and, by b), every $y\in A_{i,c}$ is visited infinitely often. Hence, $u$ is visited infinitely often. 
 Since $S$ is finite, for every path of the process there exists  a state $i$ that is visited infinitely often.
 Then, every state $j$, such that $Q_{i,u_1}Q_{u_1,u_2}\dots Q_{u_n,j}>0$ for some 
 $u_1,\dots,u_n$ and $n$,  is visited infinitely often. Since $Q$ is irreducible, for every pair $i,j$ there exist
 such $u_1,\dots,u_n$. Hence the process is recurrent. 
 \hfill $\diamond$
 
\section*{Acknowledgments}
We thank an anonymous referee for suggesting a shorter proof of Theorems \ref{th:rec} and \ref{rec2}, and all the reviewers and the Associate Editor for their thoughtful comments.

\end{document}